\documentclass[11pt]{article}
\pdfoutput=1
\usepackage{jheppub}

\usepackage[svgnames,table]{xcolor}
\usepackage{graphicx}
\usepackage{amsfonts,amsmath,amssymb,amsthm}
\usepackage{mathrsfs,bm,bbm,esint}
\usepackage[mathscr]{eucal}
\usepackage{physics}
\usepackage{slashed,soul}
\usepackage{rotating,pdflscape}
\usepackage{enumitem}
\usepackage{multirow,array}
\usepackage[numbers,sort&compress]{natbib}
\usepackage[cbgreek]{textgreek}
\usepackage{tikz}
\usetikzlibrary{cd}
\usetikzlibrary{calc}
\usetikzlibrary{decorations.pathmorphing}
\usetikzlibrary{decorations.pathreplacing}
\usetikzlibrary{decorations.markings}
\tikzset{snake it/.style={decorate, decoration=snake}}
\tikzset{->-/.style={decoration={
  markings,
  mark=at position .5 with {\arrow{>}}},postaction={decorate}}}
\tikzset{-<-/.style={decoration={
  markings,
  mark=at position .5 with {\arrow{<}}},postaction={decorate}}}
\usepackage{subcaption}
\usepackage{float}
\usepackage{afterpage}
\usepackage{cleveref}
\crefname{section}{\S\!}{\S\S\!}
\Crefname{section}{Section}{Sections}
\crefname{appendix}{Appendix}{Appendices\!}
\crefname{figure}{Fig.\!}{Figs.\!}

\definecolor{rust}{rgb}{0.8,0.2,0.2}

\newcommand{\be}{\begin{equation}}
\newcommand{\ee}{\end{equation}}
\newcommand{\bi}{\begin{itemize}}
\newcommand{\ei}{\end{itemize}}

\newcommand{\AdS}[1]{AdS$_{#1}$}
\newcommand{\lads}{\ell_\text{AdS}}

\newcommand{\bulk}{{\cal M}}
\newcommand{\bdy}{{\cal B}}
\newcommand{\bdyk}{{\sf B}^\text{k}}
\newcommand{\bdyb}{{\sf B}^\text{b}}
\newcommand{\bulkket}{{\sf M}^\text{k}}
\newcommand{\bulkbra}{{\sf M}^\text{b}}

\newcommand{\tx}{\tilde{x}}

\newcommand{\regA}{\mathcal{A}}
\newcommand{\regAc}{\mathcal{A}^c}
\newcommand{\rhoA}{\rho_{_\regA}}

\newcommand{\entsurf}{\partial\regA}
\newcommand{\Cbdy}{\Sigma_{_t}}
\newcommand{\Cbulk}{{\tilde \Sigma}_{_t}}

\newcommand{\homsurfA}{{\cal R}_{\regA}}
\newcommand{\homsurfAc}{{\cal R}_{{\regA}^c}}

\newcommand{\homsurfAcreg}{{\cal R}^\epsilon_{{\regA}^c}}
\newcommand{\homsurfAreg}{{\cal R}^\epsilon_{\regA}}

\newcommand{\fixM}{\text{\bf{\textgamma}}}

\title{Real-time gravitational replicas: Formalism and a variational principle}

\author[a]{Sean Colin-Ellerin,}
\author[b]{Xi Dong,}
\author[b]{Donald Marolf,}
\author[a]{Mukund Rangamani,}
\author[b]{Zhencheng Wang}
\affiliation[a]{Center for Quantum Mathematics and Physics (QMAP)\\
Department of Physics \& Astronomy, University of California, Davis, CA 95616, USA}
\affiliation[b]{
Department of Physics, University of California, Santa Barbara, CA 93106, USA}
\emailAdd{scolinellerin@ucdavis.edu}
\emailAdd{xidong@ucsb.edu}
\emailAdd{marolf@ucsb.edu}
\emailAdd{mukund@physics.ucdavis.edu}
\emailAdd{zhencheng@ucsb.edu}

\abstract{This work is the first step in a two-part investigation of real-time replica wormholes. Here we study the associated real-time gravitational path integral and construct the variational principle that will define its saddle-points.  We also describe the general structure of the resulting real-time replica wormhole saddles, setting the stage for construction of explicit examples.  These saddles necessarily involve complex metrics, and thus are accessed by deforming the original real contour of integration.  However, the construction of these saddles need not rely on analytic continuation, and our formulation can be used even in the presence of non-analytic boundary-sources.   Furthermore, at least for replica- and CPT-symmetric saddles we show that the metrics may be taken to be real in regions spacelike separated from a so-called `splitting surface'.  This feature is an important hallmark of unitarity in a field theory dual.
}

\begin{document}
\maketitle


\section{Introduction}
\label{sec:intro}

The path integral formulation of quantum theories has inherent advantages in Euclidean signature. In particular, the fact that the Euclidean action is non-negative in stable quantum field theories allows for a straightforward saddle point analysis. The presence of non-trivial saddles provides important insights into  non-perturbative dynamics of the theory, as illustrated by the instanton solutions in non-abelian gauge theories. A direct  first-principles Lorentzian path integral perspective on some of these issues is not as well developed for the simple reason that the rules for carrying out the stationary phase approximation have not been satisfactorily clarified.

However, despite a corresponding possible lack of rigor, path integrals may still be used to compute real-time correlation functions. To this end one often uses the Schwinger-Keldysh formalism \cite{Schwinger:1960qe,Keldysh:1964ud} (involving a mixture of Euclidean and Lorentzian path integrals) or one of its out-of-time-order generalizations \cite{Aleiner:2016eni,Haehl:2016pec,Haehl:2017qfl}.  A purely Lorentz-signature path integral may also be used if one explicitly specifies the relevant quantum state.

\begin{figure}
\centering
\begin{tikzpicture}
[scale=1.0,
tzinstate/.style={rectangle,draw=orange!90,fill=red!20,thick, inner sep=0pt,minimum size=2pt},
]
\foreach \x in {0, 4,6,8,10}
{
	\draw[tzinstate]  (\x,0)   -- ++(1,0) -- ++(-0.,0.1)  -- ++(-1,0) -- cycle;
	\draw[color=blue,thick, ->-] (\x,0)  -- ++ (0,4);
	\draw[color=blue,thick, -<-] (\x+1,0) --  ++ (0,4);
	\node at (\x+0.5,0) [below=0.5pt]{$\rho_0$};
}
\node at (0,2)  [left]{$\mathcal{U}(t;t_0)$};
\node at (1,2) [right=2pt]{$\mathcal{U}(t;t_0)^\dagger$};
\foreach\x in {5,7,9}
{
	\draw[color=blue,thick] (\x,4) -- ++ (0.5,0.5) -- ++ (0.5,-0.5);
}
\end{tikzpicture}
\caption{ A schematic illustration of the real-time (Schwinger-Keldysh) contours for the computation of the density matrix $\rho(t)$ (left) and its powers (right). The past boundary condition is supplied by the prescribed initial state $\rho_0$ and the direction of time evolution is explicitly indicated by the arrows. Forward evolution corresponds to the ket part of the state while backward evolution corresponds to the bra part.  The reduced density matrix $\rhoA(t)$ associated with some spatial domain at $t=0$ is obtained by sewing together the $\mathcal{U}(t;t_0)$ and $\mathcal{U}(t;t_0)^\dagger$ parts of the left panel along the complementary $t=0$ domain $\regAc$, while leaving open the parts along $\regA$.  It's powers thus involve similar contractions between the two parts of any given copy, while the $\regA$ parts are again contracted with neighboring copies as shown at right.
}
\label{fig:skdmat}
\end{figure}
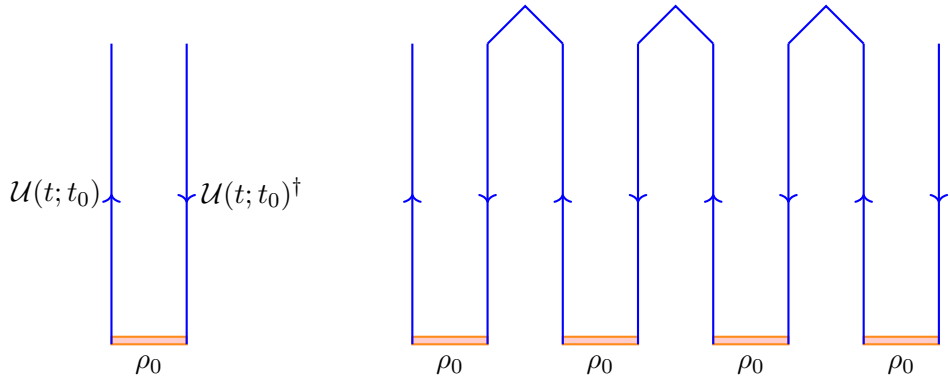

Similar prescriptions also exist for computing the R\'enyi or von Neumann entropies that capture fine-grained notions of quantum information.  To illustrate the point, consider the computation of such entropies in a quantum field theory in some time-dependent state $\rho(t)$. The state may be obtained from some past initial condition followed by evolution $\rho(t) = \mathcal{U}(t;t_0)\, \rho_0(t_0)\, \mathcal{U}(t;t_0)^\dagger$ in Lorentzian time $t$, potentially with a time-dependent Hamiltonian. The quantities of interest depend only on the information at the prescribed time $t$, and require no knowledge of the state at any future instant of time. The R\'enyi entropies which capture moments of the spectrum of the  density operator $\Tr(\rho(t)^n)$ are naturally described using an $n$-fold replication of the system along with $n$-copies each of the forward and backward time evolution. Thus we can view the quantities as being defined on a suitable timefolded contour, see \cref{fig:skdmat}.

Analogous statements also hold when computing spectral moments $\Tr(\rho_\regA^n)$ of reduced density matrices restricted to some spatial domain $\regA$ of the constant $t$ time-slice. The main difference is that this case involves a non-trivial gluing of the timefolds due to implementing the partial trace over the complement $\regAc$ of $\regA$. In the path integral, this partial trace requires sewing together along $\regAc$ the
$\mathcal{U}(t;t_0)^\dagger$ and $\mathcal{U}(t;t_0)$ parts of each copy of the time-evolved state (i.e., of each copy of the piece shown at left in \cref{fig:skdmat}).  The remaining $\regA$ regions of the $\mathcal{U}(t;t_0)^\dagger$ and $\mathcal{U}(t;t_0)$ parts of each copy are then connected cyclically as before (see again the right panel of \cref{fig:skdmat}). Readers seeking more detailed reviews may consult e.g. \cite{Dong:2016hjy,Rangamani:2016dms}.

Note that nowhere in the above discussion were we required to invoke a Euclidean formulation of the theory. However, in practice it is sometimes efficacious to perform computations entirely in Euclidean space.  This is particularly so if the quantity  is being computed at a moment of time reflection symmetry.  But analytic continuation can also be employed even in certain circumstances with  explicit time dependence.  Classic examples of the latter include computations of the growth of entanglement entropy in two dimensional CFTs following a  global  \cite{Calabrese:2005in}  or local  \cite{Calabrese:2007rg} quantum quench. In such cases, one may proceed by identifying related Euclidean configurations and computing the entropies therein as a function of Euclidean time $t_{_\text{E}}$.  At the end of the day, the result of the Euclidean computation is then analytically continued to non-trivial Lorentzian times $t$ by Wick rotating $t_{_\text{E}} \to i\, t$. This allows one to obtain the desired real-time evolution of the entropies.

The reasons for employing the Euclidean crutch are two-fold. Firstly, at a technical level one can exploit the fact that the replication is geometrically easy to understand as the construction of a branched cover geometry, and in particular one that is branched over a codimension-2 locus.  The spectral moments of the density matrix are then given by partition functions on the branched cover. Secondly,  one can evaluate such partition functions in a semiclassical approximation using a saddle-point analysis which, as for the instantons, is facilitated in the field theory context by the boundedness of the Euclidean action. A key fact that will play a crucial role below is that the resulting entropies  are manifestly real in this
computational scheme since they are written in terms of real Euclidean configurations.

The Euclidean perspective can thus be an efficient technical tool to eke out the answer for real-time  quantities of interest.  Unfortunately,  it also has some major disadvantages.  The first is that it is not obviously useful in contexts with general smooth-but-nonanalytic sources.  And for related reasons, it can be difficult to use when Lorentzian sources are only approximately known (perhaps because they have been computed numerically). But from the conceptual point of view the main issue with a Euclidean approach is that the real-time dynamics lies obscured. While this lacuna is not particular to entropic quantities --
the real-time dynamics of instantons is also not well understood -- it ends up being striking in this context, especially when we also allow gravity  to be dynamical, as for instance in the duals of holographic field theories.

To illustrate this, recall that holographic entanglement entropy  prescriptions \cite{Ryu:2006ef,Hubeny:2007xt} posit that the von Neumann entropy is computed by the area of a bulk codimension-2 extremal surface at leading (planar) order  in an expansion at large central charge $c_\text{eff}$, or equivalently at small bulk Newton constant $G_N$ with the bulk AdS scale $\lads$ held fixed.\footnote{ It is convenient to define $c_\text{eff}  = \frac{\lads^{d-1}}{16\pi G_N}$ as the effective central charge of a CFT$_d$ dual to gravity in \AdS{d+1}.\label{fn:ccharge}} The time-reflection symmetric RT prescription \cite{Ryu:2006ef} is justified by a gravitational version of the replica construction  \cite{Lewkowycz:2013nqa} (following earlier such constructions \cite{Fursaev:2006ih,Headrick:2010zt,Casini:2011kv,Hartman:2013mia,Faulkner:2013yia}). The construction proceeds by noting that the standard AdS/CFT dictionary requires the dual gravitational spacetime to be the lowest action saddle-point solution to the Euclidean quantum gravity path integral whose boundary is the branched cover geometry described earlier.\footnote{  Technically, one has to worry about the fact that in gravitational theories (in contrast to quantum field theories) the Euclidean action is of indefinite sign owing to the conformal mode. The standard prescription \cite{Gibbons:1978ac}  is to analytically continue the conformal mode to a presumed steepest-descent contour (which involves purely imaginary scale factors at $2^\text{nd}$ order about familiar solutions). We will assume that such a prescription has been employed to render the semiclassical Euclidean quantum gravity path integral sensible.} The analysis of \cite{Lewkowycz:2013nqa}, which at its core is a generalization of  the Gibbons-Hawking prescription for computing partition functions in Euclidean quantum gravity \cite{Gibbons:1976ue}, focused on recovering the von Neumann entropy by noting that the gravitational replica construction takes a particularly simple form when analytically continued to replica numbers $n$ close to $1$. But this prescription also helps one understand the gravity dual of R\'enyi entropy \cite{Dong:2016fnf}.

This Euclidean quantum gravity perspective has been enormously helpful in clarifying various aspects of the holographic entanglement. For instance, it was used to establish that subleading corrections are given in terms of bulk  entanglement across the homology surface\footnote{ The homology surface is a partial Cauchy surface of the bulk geometry whose boundaries are the extremal surface and the subregion of interest on the asymptotic boundary of the spacetime. The entanglement wedge is the bulk domain of dependence of this region cf.,   \cite{Headrick:2014cta}.  The notion of homology surface was originally introduced in \cite{Headrick:2007km} in contexts restricted to time-reflection symmetry. } \cite{Faulkner:2013ana}. More generally, the quantum extremal surface prescription of \cite{Engelhardt:2014gca} argues that, to all orders in a perturbative expansion at large central charge, the von Neumann entropy is given by extremizing the generalized entropy which is the combination of the leading order classical area term and the subleading bulk entanglement entropy. And this stronger result is likewise justified by an analogous variational argument in the Euclidean path integral \cite{Dong:2017xht,Penington:2019kki,Almheiri:2019qdq}.

While the standard derivation of the quantum extremal surface proposal stems from the Euclidean path integral, its physical implications are most striking in the realm of real-time evolution in the context of the black hole information paradox. As argued originally by \cite{Penington:2019npb,Almheiri:2019psf} the quantum extremal surface associated with the entire boundary in an evaporating black hole undergoes a dynamical phase transition after which the entanglement wedge of the boundary no longer includes a portion of the black hole interior (called a quantum extremal `island' in \cite{Almheiri:2019hni,Almheiri:2019yqk}). This in particular reproduces the expected Page curve \cite{Page:1993df} of the evaporating black hole. A summary of these developments from the past year can be found in the review  \cite{Almheiri:2020cfm}.

\begin{figure}[h]
\centering
\begin{tikzpicture}[scale=0.6]
\coordinate (a) at (0,-4);
\coordinate (b)  at (-5,3);
\coordinate (c)  at (5,3);
\draw[black,fill=gray!20,opacity=0.8] (a) ellipse (2 and 1.5);
\draw[blue,fill=blue!30] (a) ellipse (0.8 and 0.4);
\draw[red,thick]  ($(a) + (-0.8,0)$) .. controls (-1,0) and (-2.5,2) .. ($ (b) +(-0.4,-0.6)$);
\draw[red,thick] ($(a) + (0.8,0)$) .. controls (1,0) and (2.5,2)..  ($ (c) +(0.4,-0.6)$);
\draw[red,thick]  ($(b) +(0.4,0.6)$) .. controls (-2,1.5) and (2,1.5).. ($ (c) +(-0.4,0.6)$);
\draw[rotate around={60:(b)},black,fill=gray!20,opacity=0.8] (b) ellipse (2 and 1.5);
\draw[rotate around={60:(b)},blue,fill=blue!30] (b) ellipse (0.8 and 0.4);
\draw[rotate around={-60:(c)},black,fill=gray!20,opacity=0.8] (c) ellipse (2 and 1.5);
\draw[rotate around={-60:(c)},blue,fill=blue!30] (c) ellipse (0.8 and 0.4);
\draw[thick,red,fill=red!20,opacity=0.7] (0,-0.5) ellipse (1.5 and 0.35);
\end{tikzpicture}
\caption{A  replica wormhole spacetime that contributes in to the Euclidean path integral for the computation of spectral traces of density matrices.  }
\label{fig:EreplicaWH}
\end{figure}
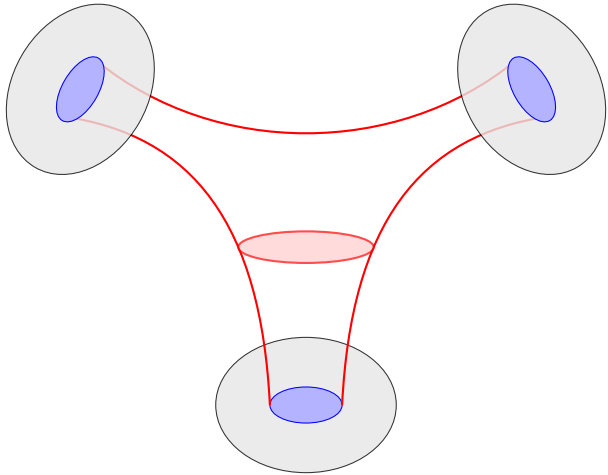

As described above, the justification in most of the literature for the use of these quantum extremal surfaces has been based on Euclidean gravitational replica saddle-point geometries. Given the boundary replica spacetime, the presence of dynamical gravity in the bulk allows non-trivial replica symmetric configurations that connect different replica copies, leading to replica wormholes \cite{Penington:2019kki,Almheiri:2019qdq}; see the schematic illustration in \cref{fig:EreplicaWH}. Such configurations are allowed because the quantum gravity path integral instructs us to sum over all geometries which can connect the given boundaries.  In general there may be several different ways to do so respecting the boundary conditions and symmetries, perhaps allowing several different topologies for the Euclidean bulk. These geometries can, and often do, exchange dominance as one varies parameters. In particular, \cite{Penington:2019kki,Almheiri:2019qdq} argued that this  leads to the aforementioned phase transition associated with the Page curve.

While replica wormholes define sensible Euclidean manifolds, the real-time dynamics of the above phase transition remains obscure. And while the technical construction in \cite{Penington:2019kki,Almheiri:2019qdq} exploited specific features of low dimensional gravity models, as emphasized in \cite{Marolf:2020xie} the basic lesson  is that one should expect non-trivial replica wormhole contributions to general Euclidean path integrals.\footnote{ Although we have couched our discussion in the holographic setting (in which the bulk spacetime is asymptotically AdS), the democratic nature of gravity makes it clear that similar statements apply in other contexts. See \cite{Gautason:2020tmk,Anegawa:2020ezn,Hashimoto:2020cas,Hartman:2020swn,Dong:2020uxp,Chen:2020tes,Krishnan:2020fer,Hartman:2020khs,Marolf:2020rpm}  for analyses of replica wormholes with other asymptotic boundary conditions.}

These developments thus suggest an important link between gravitational entropy and spacetime wormholes. But while this connection has been well fleshed out in the Euclidean context,  the corresponding real-time story remains more obscure. Indeed, it is a classic result   that replica-wormhole-like topologies do not admit smooth Lorentz-signature metrics. This is particularly clear if we assume replica symmetry, which for an $n$-fold replica would  require the replica-invariant surface to have $2n$ distinct timelike normals with the property that no such normal can be deformed to any other while remaining timelike.  Such surfaces cannot occur in smooth Lorentz-signature manifolds when $n \ge 2$.  So one may ask what form any corresponding real-time saddles might take.  Since black hole evaporation involves dynamical time-dependent geometries, a clearer understanding of the real-time description is imperative if we are to fully fathom the impact of spacetime wormholes on black hole information.

With these motivations in mind, let us revisit the real-time setup for computing entropies in gravitational theories.\footnote{As argued in \cite{Marolf:2020rpm}, this prescription is properly taken to compute `swap entropies.'} Given a timefold replica contour at the boundary, our task is to ascertain the stationary points of the bulk Lorentzian Einstein-Hilbert action.\footnote{ While the discussion here can be generalized to other gravitational theories to include higher derivative couplings or matter fields, for the sake of simplicity we will focus on just Einstein-Hilbert dynamics.} In the AdS/CFT context this problem was examined in \cite{Dong:2016hjy} for use in deriving the covariant holographic entanglement entropy proposal.  The authors of that work used the real-time AdS/CFT prescription of \cite{Skenderis:2008dh,Skenderis:2008dg} to motivate the resulting bulk spacetimes, primarily focusing on the von Neumann entropy ($n\to 1$). In the latter limit one has the advantage that the field equations may be analyzed perturbatively in the replica parameter $(n-1)$, which allows for considerable simplification. While the construction is in no way restricted to this regime, the analysis for $n>1$ was not hitherto carried out in detail (though see Appendix A of \cite{Dong:2016hjy}).

We will undertake a careful examination below of real-time gravitational replica saddles at finite $n-1$, arguing that the Einstein-Hilbert action does in fact admit stationary points but that the associated spacetimes have complex-valued metrics. In particular, in our formalism the time coordinate will remain real but the metric will take complex values in the spacetime interior (though boundary conditions will require the induced metric to remain real on e.g. timelike asymptotically-AdS boundaries). In simple cases the complex nature of the bulk can be encapsulated in an appropriate $i\varepsilon$ condition.

The appearance of complex solutions should come as no surprise.
From the Schwinger-Keldysh point of view, it is natural to construct the quantum state $\rho_0$ of \cref{fig:skdmat} from an imaginary-time path integral, so that real spacetimes cannot describe the full timefolded field theory contour.  More generally, while one may expect the real-time equations of motion governing stationary points to be real, there is no reason to expect this for the boundary conditions imposed by attaching any particular quantum state.  Indeed, vacuum states for harmonic oscillators famously tend to impose positive-frequency boundary conditions which render any associated solutions intrinsically complex.\footnote{ Real-time correlation functions in thermal states of holographic field theories have also been argued to be computed on complex gravitational Schwinger-Keldysh geometries \cite{Glorioso:2018mmw}. But while detailed analysis of observables shows consistency  with field theory expectations \cite{Chakrabarty:2019aeu,Jana:2020vyx}, a first-principles argument that they are unique saddle point of the gravitational path integral does not yet exist.  Nevertheless, we will argue in \cref{app:thermal} that these geometries do provide an useful arena to illustrate some of the ideas we discuss. \label{fn:skgrav}}  See e.g., section 4.3 of \cite{Marolf:2004fy} for a discussion in language closely connected to that used here.  In particular, it is the complex nature of such real-time stationary points that allows them to contribute anything other than a pure phase, and thus to  reproduce Euclidean results in appropriate contexts.\footnote{In many cases this again leads to an $i\varepsilon$ prescription.  Thus the above comment is then largely equivalent to noting that one might regularize the sewing of bra spacetimes to ket spacetimes by including a small Euclidean piece of spacetime, and that doing so would then require excursions into complex time.  This one again suggests that the final solution is complex.} Indeed, any stationary point that is invariant under the natural CPT-conjugation symmetry of any R\'enyi calculation (associated with exchanging the bra and ket parts in \cref{fig:skdmat}) must contribute a real {\it amplitude} $e^{iS_L}$, thus necessitating an imaginary Lorentzian action $S_L$.  This is in fact a manifestation of the general phenomenon noted some time ago \cite{Louko:1995jw} that singularities in the Lorentz-signature causal structure are associated with imaginary contributions to the Lorentzian action.

The goal of this work is to set up the variational problem for the determination of the gravitational replica saddle which computes the $n^{\rm th}$ (swap) R\'enyi entropy of an appropriate density matrix $\rho_0$.  This sets the stage for constructing examples of such saddles in the companion paper  \cite{Colin-Ellerin:2020exa}. Lorentz-signature replica wormholes and the associated Lorentzian path integrals were also recently discussed in \cite{Marolf:2020rpm}, though the form of the saddle-point geometries was not analyzed in detail.

We begin in \cref{sec:skreplicas} with a discussion of the relevant Lorentzian path integrals, emphasizing the space of bulk configurations over which we choose to sum.  With replica boundary conditions, the bulk configurations are naturally called Lorentz-signature replica wormholes.  We then set up an associated variational problem in \cref{sec:varprob}.  In particular, the Einstein-Hilbert action extends naturally to our somewhat-singular Lorentzian configurations using ideas from \cite{Louko:1995jw} associated with the complex generalization of the Gauss-Bonnet theorem.   And borrowing from the Euclidean discussion of \cite{Dong:2019piw} allows us to introduce boundary conditions that make the associated variational principle well-defined.   In the process we will elaborate on the symmetries of the replicas and  general expectations for the saddle-point spacetimes.  For completeness, we also present a brief discussion of initial conditions for our path integrals in \cref{sec:stateprep}.  Elements of these discussions are already present in \cite{Dong:2016hjy} (see also \cite{Marolf:2020rpm}); our aim here is to elaborate on certain aspects for clarity, and to describe in detail the real-time variational problem for replica wormhole saddles.   We close with a summary and discussion of future directions in \cref{sec:Disc}.   The appendices contain some additional details of corner term contributions and an example with smooth complex valued metrics.

\section{Path integrals for the density matrix and replicas}
\label{sec:dmatrep}
\label{sec:skreplicas}

As described above, we would like to  compute entropies (or, more precisely, swap entropies \cite{Marolf:2020rpm}) in holographic field theories without employing analytic continuation.  To do so, let us first discuss the situation for standard quantum mechanical theories on a fixed spacetime (and thus without dynamical gravity) in \cref{sec:RPISQT}.  We then address the case of dynamical gravity in \cref{sec:RPIDG}.

\subsection{Replica path integrals in standard quantum theories}
\label{sec:RPISQT}

We begin by discussing the path integral that computes matrix elements of $\rho(t)$.
Given an initial state $\rho_0$ at time $t_0$,\footnote{We will discuss initial states in \cref{sec:stateprep} below.} we can construct the density matrix $\rho(t)$ on a Cauchy slice $\Cbdy$ by applying time-dependent Hamiltonian evolution to write $\rho(t) = \mathcal{U}(t; t_0) \, \rho_0\, \mathcal{U}(t;t_0)^\dagger$.
Taking a partial trace $\rhoA(t) = \Tr_{\regAc}(\rho(t))$ then gives its restriction to a subregion $\rhoA(t)$, where $\regA$ and $\regAc$ are complementary to each other at time $t$.  Here $ \mathcal{U}(t; t_0)$ includes any explicit sources that may be introduced in the course of the evolution.  Note that the path integral constructing the elements of the density matrix $\rho$ involves a piece implementing $\mathcal{U}(t; t_0)$ for the  forward evolution of its `ket' $\ket{\psi}$ as well as another implementing $\mathcal{U}(t; t_0)^\dagger$ for the backward evolution for its `bra'  $\bra{\psi}$. One may think of the former as integrating over fields in a `ket spacetime' and the latter as integrating over fields in a `bra spacetime,' with configurations weighted by $e^{i(S^\text{k} - S^\text{b})}$, with $S^\text{k},S^\text{b}$ being the standard actions for fields on the ket and bra spacetimes respectively.

\begin{figure}[htbp]
\begin{center}
\includegraphics[width=3in]{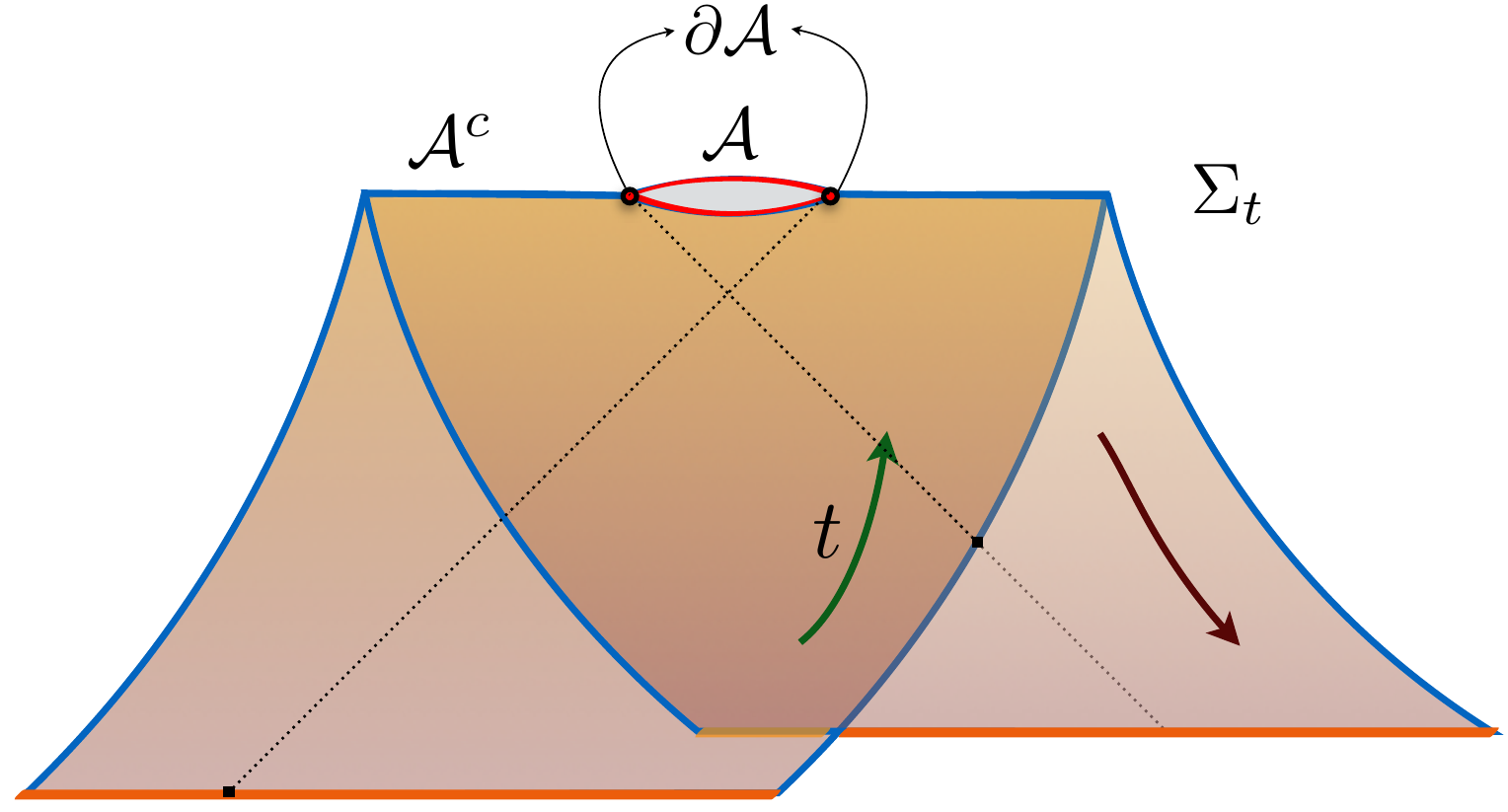}
\caption{ The timefolded Schwinger-Keldysh geometry $\bdy$ computing the matrix elements of $\rhoA$ for a quantum theory on a fixed background. The forward and backward evolutions are glued together on the partial Cauchy slice $\Cbdy$, except for a cut around $\regA$. We have also indicated the past light-cones from the entangling surfaces which serve to demarcate the Milne and the Rindler wedges defined in the main text. In the gravitational context, the boundary geometries will be of this form. }
\label{fig:bdyrho}
\end{center}
\end{figure}

To construct the reduced density matrix $\rhoA$, one traces $\rho(t)$ over the Hilbert space associated with the region $\regAc$.  This is implemented in the path integral by identifying the associated ket and bra spacetimes at the final time along $\regAc$.\footnote{ One can equivalently project $\rho$ against the maximally entangled state supported on two copies of $\regAc$, though this requires a notion of CPT conjugation to turn kets into bras.} The result of this sewing is a time-folded (Schwinger-Keldysh) spacetime $\bdy$ over which the path integral is to be performed, see \cref{fig:bdyrho}.

The time-folded spacetime $\bdy$ has  three causal domains of interest. Focusing on the ket piece of the contour we have:
\begin{enumerate}[label=\roman*.,leftmargin=0pt]
\item The set $J^-[\entsurf]$  describing the causal past of $\entsurf$.  We will refer to  $J^-[\entsurf]$ as the \emph{Milne wedge}.
\item The past domain of dependence $D^-[\regA]$ of $\regA$. We will refer to this region as  the \emph{Rindler wedge} of $\regA$.
\item The analogous Rindler wedge of $\regAc$ defined as past domain of dependence $D^-[\regAc]$ of $\regAc$.
\end{enumerate}
These regions are separated by `Rindler horizons' defined by the past-directed null congruences orthogonal to $\entsurf$. The same regions are present on the bra piece of the contour.  In particular, note that the causal nature of the Schwinger-Keldysh construction ensures that we consider only regions to the {\it past} of $\regA$ and $\regAc$ on both bra and ket pieces of the contour, since we have not evolved the system beyond the Cauchy surface at time $\Sigma_t$.

\subsection{Replica path integrals with dynamical gravity}
\label{sec:RPIDG}

We now discuss contexts with dynamical gravity, following the same basic approach as in \cref{sec:RPISQT}.  For simplicity, we assume the system to be asymptotically AdS, and thus to have a well-defined notion of a boundary spacetime (at least in some given conformal frame).  It is also convenient to assume some notion of AdS/CFT duality (though perhaps involving an ensemble of dual field theories as in \cite{Saad:2019lba}, and as suggested by general discussions in \cite{Coleman:1988cy,Giddings:1988cx,Giddings:1988wv,Marolf:2020xie,Marolf:2020rpm}) so that we can motivate our bulk construction using the dual CFT.

\begin{figure}[htbp]
\begin{center}
\includegraphics[width=2in]{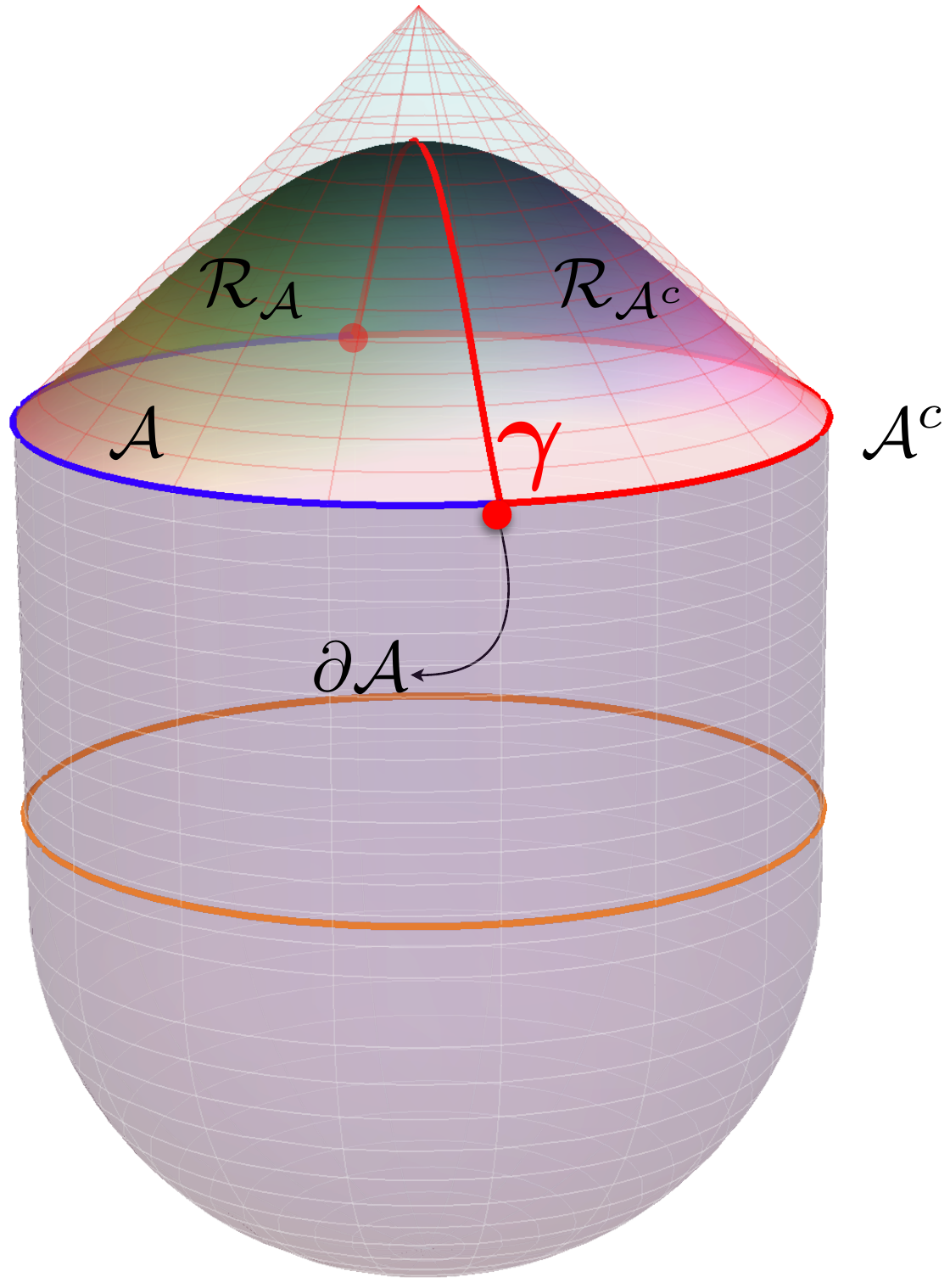}
\caption{The bulk domains of interest in the Lorentzian construction for either the ket or the bra spacetime. Given a partition of $\Cbdy$ into regions $\regA$ and $\regAc$, any bulk Cauchy surface $\tilde{\Sigma}_t$  with $\partial \Cbulk = \Cbdy$ admits a decomposition $\tilde{\Sigma}_t = \homsurfA \cup \homsurfAc$. These domains are separated by a bulk codimension-2 surface $\fixM$, which is anchored on the entangling surface. On saddle point solutions this surface approaches the extremal surface as $n\to 1$.}
\label{fig:bulkDoms}
\end{center}
\end{figure}

In particular, we first describe a bulk path integral that may be said to compute the bulk description of matrix elements of the dual field theory restricted density matrix $\rho_\regA (t)$.
The boundary conditions for this path integral will be defined by the Schwinger-Keldysh contour $\bdy$ associated with matrix elements of $\rho_\regA (t)$ described in \cref{sec:RPISQT} above.  Here $\regA$, $\regAc$ are complementary regions within some boundary Cauchy slice $\Sigma_t$.

We will also need some further elements to describe the bulk geometries $\bulk$ (with $\partial \bulk = \bdy$) over which our path integral will sum. As in the boundary, the bulk spacetimes $\bulk$ will consist of bra and ket parts.  Focusing first on the ket part, we imagine that we can use a standard bulk path integral to evolve the initial state wavefunction of the bulk forward (in the usual ADM sense) up to a bulk Cauchy slice $\Cbulk$ with $\partial \Cbulk =\Cbdy$. The slice $\Cbulk$ is not uniquely determined given $\Cbdy$, but since Cauchy surfaces are achronal the surface $\Cbulk$ must  be everywhere spacelike separated from $\Cbdy$; see \cref{fig:bulkDoms}. At an extreme, we can take $\Cbulk$ to be  null, straddling the past boundary of the bulk domain of influence of $\Cbdy$.

We now partition $\Cbulk$ into two regions by introducing a bulk codimension-2 surface $\fixM$ lying in $\Cbulk$ (and thus which is codimension-1 in $\Cbulk$).  Following \cite{Marolf:2020rpm}, we will  call $\fixM$ the splitting surface, though we will sometimes also refer to $\fixM$ as the cosmic brane.\footnote{It is perhaps natural to reserve the term `cosmic brane' for the case where $\fixM$ is associated with a non-trivial delta-function in the Ricci scalar while using `splitting surface' to include the case where the coefficient of this delta-function vanishes.}  The location of $\fixM$ in a saddle-point geometry may eventually be determined dynamically from the variational principle, though for the moment we simply sum over all possible choices of $\fixM$ (and also over all inequivalent choices of $\Cbulk$, see further discussion below).   The two resulting parts of $\Cbulk$ will be called the homology surfaces $\homsurfA$ and $\homsurfAc$ respectively of $\regA$ and $\regAc$, with $\partial \homsurfA = \regA \cup \fixM$ and
$\partial \homsurfAc = \regAc \cup \fixM$.  This partition will also endow the bulk with three distinguished causal domains: the causal past $\tilde{J}^-[\fixM]$ of the separatrix cosmic brane $\fixM$, and the  past domains of dependence $\tilde{D}^-[\homsurfA]$ and $\tilde{D}^-[\homsurfAc]$ of the two homology surfaces. We will refer to the latter domains as the past homology wedges. We illustrate these bulk regions in \cref{fig:bulkDoms}.

\begin{figure}[h!]
\begin{center}
\includegraphics[scale=0.4]{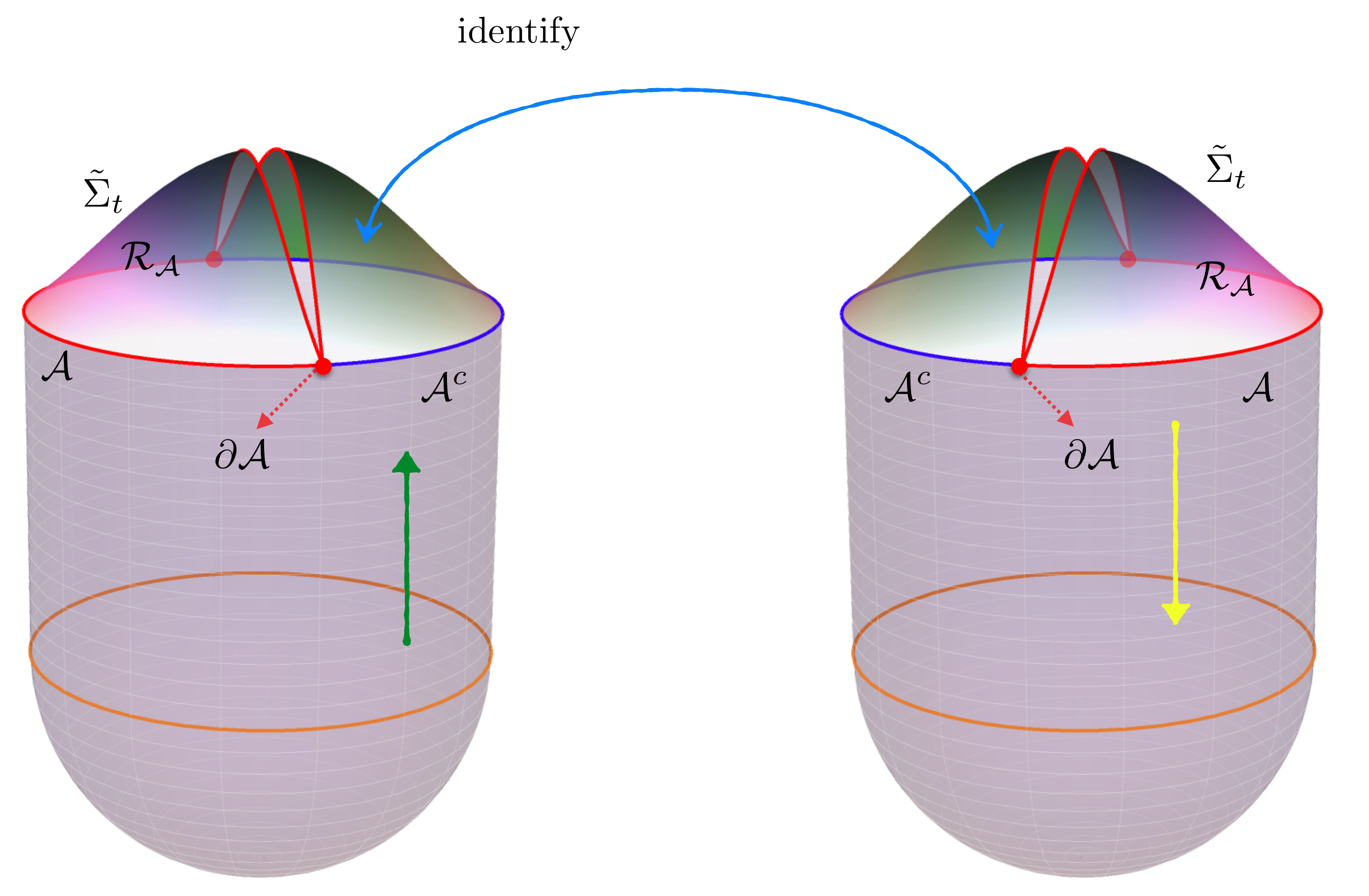}
\caption{Bulk configurations relevant to computing the bulk dual of the boundary density matrix  $\rhoA(t)$. The forward evolution for $\ket{\Psi}$ (left) proceeds up to
$\Cbulk$, while the backwards evolution for  $\bra{\Psi}$ starts there (right). We have identified the bra and ket spacetime along $\homsurfAc$ in accord with the prescription for the bulk dual of $\rhoA(t)$. Further gluing the two spacetimes together along $\homsurfA$ and summing over all metrics would compute the trace of $\rhoA(t)$. }
\label{fig:skrho1}
\end{center}
\end{figure}

The bra part of the geometry is constructed similarly, except that one now evolves back to the initial state wavefunction. At the AdS boundary, the gluing between the bra and ket spacetimes is determined by the regions $\regA$ and $\regAc$ on the boundary. It is natural to extend this gluing into the bulk by identifying the bra and ket spacetimes along $\homsurfAc$ and summing over bulk geometries to obtain the bulk dual of the reduced density matrix $\rhoA$.  Note that this sum over geometries can be said to  implicitly sum over all inequivalent choices of $\Cbulk$ and $\fixM$, or alternatively to implicitly sum over all inequivalent choices of the resulting $\homsurfA$ and $\homsurfAc$.

In the above,  we have basically chosen the bulk configurations over which we sum to mimic the boundary contour $\bdy$, replacing the sewing across the boundary region $\regAc$ with sewing across the homology surface $\homsurfAc$.  One can attempt to motivate this in a neighborhood of the boundary by the usual Fefferman-Graham expansion.   And one can further motivate this prescription by cutting open bulk path integral computations of traces of powers of the density matrix in familiar cases as in \cite{Dong:2018seb} (in the discussion surrounding its figure 3). But for the present work we will simply declare the above to be our recipe, leaving for the future any attempt to upgrade the above naturalness arguments into a complete derivation.  This will then lead to an ansatz for a bulk path integral that compute boundary (swap) R\'enyi entropies by summing over what we call Lorentz-signature replica wormholes.  As we show in \cref{sec:varprob}, the key point will then be that such replica wormholes lead to a well-defined Einstein-Hilbert variational problem.  Furthermore, when appropriately complexified, such real-time replica wormholes can provide stationary points that allow our path integral to be studied semiclassically.

To summarize, bulk configurations of the path integral associated with matrix elements of $\rhoA$ have the following ingredients:
\begin{itemize}[wide,left=0pt]
\item An initial state wavefunction  (or an Euclidean end-cap geometry, see \cref{sec:stateprep}), prescribing the state $\rho_0$ from which we evolve.
\item Lorentzian sections of the geometry for the ket and bra parts, with the flow of time dictated by the evolution direction specified at the boundary.
\item A gluing condition across a homology surface $\homsurfAc$, with the flow of time reversing as we cross the gluing surface.
\end{itemize}

We now have the ingredients in place to set up the replica computation and define the bulk path integral for $\Tr(\rhoA^n)$.  To define the boundary conditions for the bulk path integral, we begin with $n$-copies of the bra and ket boundary geometries, each constructed in \cref{sec:RPISQT} above. Labeling the boundary ket spacetimes as $\bdyk_i$ and the boundary bra spacetimes as $\bdyb_i$ (with $i=1,2,\cdots, n$), we sew these geometries into a R\'enyi boundary spacetime $\bdy_n$ by gluing $\bdyk_i$ onto $\bdyb_i$ along $\regAc$, and onto $\bdyb_{i-1}$ along $\regA$.  Here additions involving the index $i$ are performed modulo $n$.

\begin{figure}[h!]
\begin{center}
\includegraphics[scale=0.4]{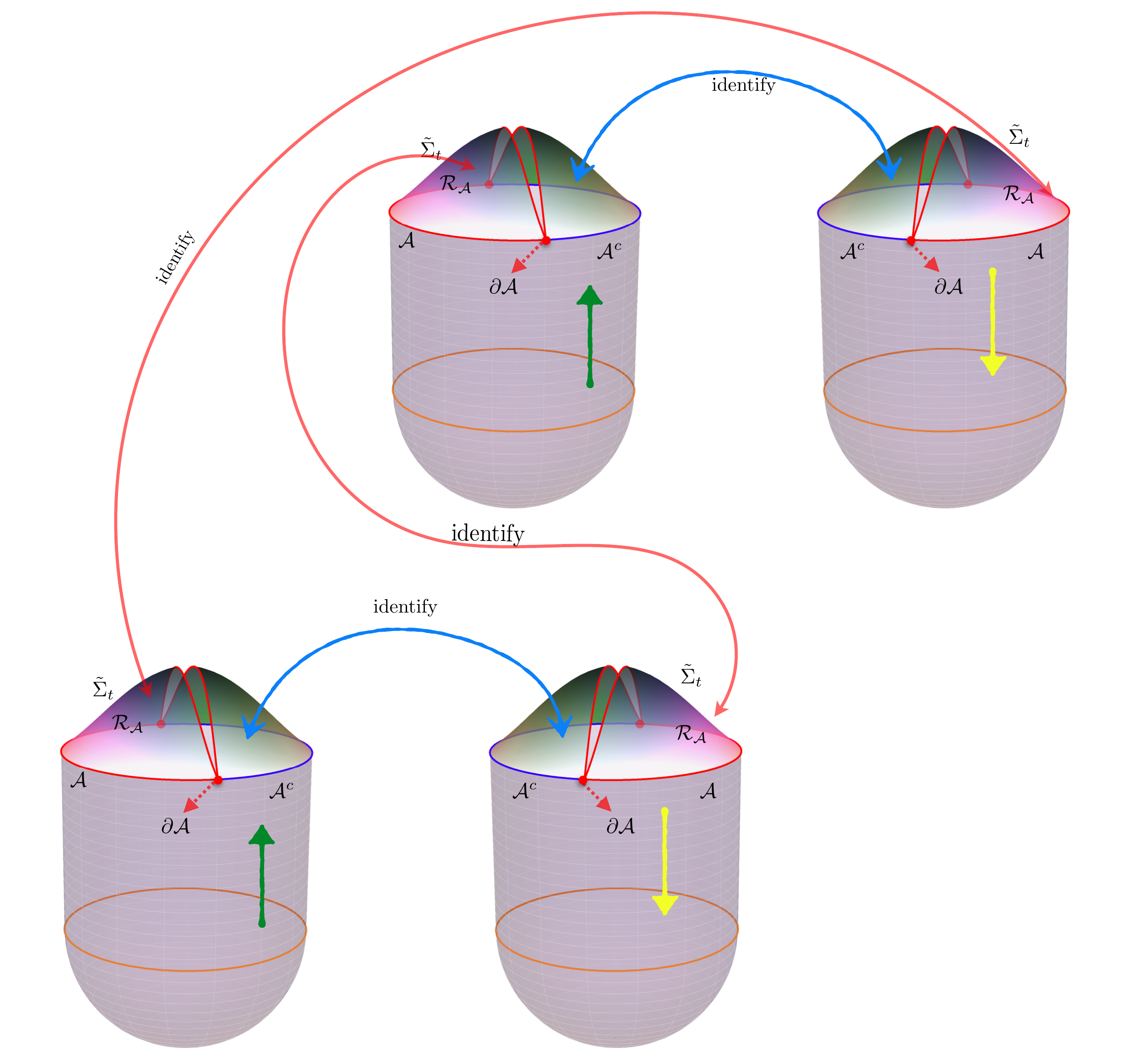}
\caption{The form of configurations over which we sum in the bulk path integral for $\Tr(\rhoA^2(t))$.  We again use the conventions described in \cref{fig:skrho1}. }
\label{fig:skrho2}
\end{center}
\end{figure}

Turning now to the bulk, each path integral configuration will be formed from $n$ bulk  ket pieces $\bulkket_i$ and $n$ bulk bra pieces $\bulkbra_i$, with $i=1,2,\cdots, n$.  We will sew these geometries into a bulk R\'enyi configuration by gluing $\bulkket_i$ onto $\bulkbra_i$ along $\homsurfAc$, and onto $\bulkbra_{i-1}$ along $\homsurfA$.  As in \cite{Marolf:2020rpm}, we use the term Lorentz-signature replica wormholes to refer to any bulk spacetime of the above form, regardless of whether or not it is a stationary point of a variational principle.
We depict such a configuration for $n=2$ in
\cref{fig:skrho2}.

Note that above-described space of configurations for the bulk replica path integral has the following discrete symmetries:
\begin{itemize}
\item A cyclic $\mathbb{Z}_n$ replica symmetry that acts by shifting the bra and ket spacetimes by  $i \to i+1, \mod(n)$.
\item For each copy $i$ of the density matrix we have a $\mathbb{Z}_2$ involution associated with the CPT map (which involves anti-linear complex conjugation).  This map exchanges corresponding bras and kets, with the $i^\text{th}$ such CPT map acting as $\bulkket_{j} \leftrightarrow \bulkbra_{i-j}$ (using addition modulo $n$).
\end{itemize}

By definition, our path integral sums over real such Lorentz-signature replica wormholes.  However, in looking for saddle points $\bulk_n$ we will wish to deform the contour of integration to allow complex metrics.  Indeed, it is familiar from standard quantum mechanics that processes forbidden by standard classical evolution are dominated by a complex saddle-point, as in the use of Euclidean paths to compute quantum tunneling.  And in the gravitational R\'enyi context, it is clear that smooth Lorentz-signature metrics cannot yield saddles since for $n> 1$ the causal structure at $\fixM$ is ill-defined.  But we may hope to find complex saddles when the time coordinate remains real by allowing the metric to be complex.

It will thus be of interest below to study the action of the above symmetries on complex saddles.
In doing so, in order to preserve the reality of $ \Tr( \rho^n)$, the CPT maps should be extended so that they complex-conjugate the metric as well.  Note that complex spacetimes which are symmetric under all of these symmetries remain real on the gluing surfaces (i.e., along $\homsurfA$ and $\homsurfAc$ on each of the $2n$ copies of the spacetime), and this will in particular be true of symmetric saddles.\footnote{ We will focus below on the solutions that preserve replica symmetry: it would be interesting to understand whether solutions that break the aforementioned symmetries could play a role analogous to replica symmetry breaking saddles; see e.g., \cite{Penington:2019kki,Dong:2020iod,Marolf:2020vsi,Akers:2020pmf}.} Furthermore, we expect deformations (of the location)  of the homology surfaces $\homsurfA$ and $\homsurfAc$ that do not affect $\fixM$ to describe unitary evolution of the bulk that will cancel exactly between the bra and ket spacetimes on either side of the surface.  We thus correspondingly expect that while the saddle-point dynamics may in some sense determine $\fixM$,  it should not lead to preferred choices for $\homsurfA, \homsurfAc$.  Symmetric saddle-points should thus remain real and Lorentz-signature within the entire homology wedges of $\homsurfA, \homsurfAc$ so that the actions for these regions cancel between the bra and ket spacetimes for all choices of $\homsurfA, \homsurfAc$.  We will justify this expectation more fully when we study the variational principle below, but the upshot is that for symmetric saddles the homology wedges $\tilde{D}^-[\homsurfA]$ and $\tilde{D}^-[\homsurfAc]$ will remain well-defined, and that such saddles will be complex only outside these homology wedges (in a region that we may still call the past Milne wedge $\tilde{J}^-[\fixM]$ of $\fixM$).

\section{The variational problem}
\label{sec:varprob}
\Cref{sec:dmatrep}  described the branched and time-folded bulk spacetimes over which our replica path integral will sum, but we have not yet addressed the bulk dynamics in detail.  Of course, we wish to take this to  be described by the Einstein-Hilbert action.  But while this action is familiar when evaluated on smooth Lorentz-signature spacetimes, it may not be immediately clear how to define contributions to this action from the region near $\fixM$.  Furthermore, the choice of action is intimately tied to the construction of a good variational problem.  In particular, to ensure a good semiclassical limit we must show that varying our action within the class of allowed variations leads precisely to an appropriate formulation of the Einstein equations (without extra constraints).

We address these issues below.  We begin in \cref{sec:varaway} by quickly discussing the spacetime away from $\fixM$.  We then define contributions to our Einstein-Hilbert action from the region near $\fixM$ in \cref{sec:cover} and discuss boundary conditions to be imposed at $\fixM$ in sections \cref{sec:imtime} and \cref{sec:realtime}.  Together, these yield a well-defined variational principle as desired.  All of this treats the full $n$-fold replica geometry $\bulk_n$ with boundary $\bdy_n$.  \cref{sec:fundomain} then briefly describes the implications when one imposes replica symmetry and describes the bulk using the geometry $\widehat{\bulk}_n = \bulk_n/{\mathbb Z}_n$ associated with a single fundamental domain.

\subsection{The action and variations away from the splitting surface}
\label{sec:varaway}

Heuristically, we define the gravitational dynamics by the real-time path integral:
\begin{equation}\label{eq:skMn}
Z[\bdy_n] := \int_n [Dg] \, e^{i\, S},
\end{equation}	
with the subscript $n$ reminding us that we are computing the path integral over bulk replica wormholes with the gluing conditions specified above and with boundary conditions defined by $\bdy_n$. We will take
\begin{equation}
\label{eq:TotalS}
S = S_\text{gr}^k - S_\text{gr}^b + S_{\fixM},
\end{equation}
where $S_\text{gr}^k$ is the standard gravitational action for the ket  parts of the spacetime, $S_\text{gr}^b$ is the standard gravitational action for the bra parts of the spacetime, and $S_{\fixM}$ is a contribution to $S$ from the splitting surface $\fixM$ that we will address more carefully below.
If we can define and solve the variational problem for the bulk geometry $\bulk_n$, the R\'enyi entropies will be obtained from  the standard formula
\begin{equation}\label{eq:renyiMn}
S^{(n)}_\regA = \frac{1}{1-n}\, \log\left(\frac{Z[\bdy_n]}{Z[\bdy]^n}\right)  = \frac{1}{n-1} \left(I_n - n \, I_1 \right)\, ,
\end{equation}	
where $I_n :=  -\log Z[\bdy_n]  \approx- i\, S[\bulk_n]$.

The gravitational action with which we work is the Einstein-Hilbert action, supplemented with the usual Gibbons-Hawking term  at any boundaries.  We also include additional counter-terms $S_\text{ct}$ at the asymptotic boundary as required.   For any bra or ket piece ${\cal M}$ of ${\cal M}_n$ we thus have
\begin{equation}\label{eq:EHact}
\begin{split}
S_\text{gr}[\bulk] &=
	 \frac{1}{16\pi G_N} \,  \int_{\bulk}\, d^{d+1} x\, \sqrt{-g} \left[ R + \frac{d(d-1)}{\lads^2} \right] + \frac{1}{8\pi G_N} \int_{\bdy}\, d^{d} x\, \sqrt{|\gamma|} \, K  \\
& \qquad \qquad
	+ \frac{1}{8\pi G_N} \int_{\homsurfA \cup \homsurfAc}\, d^{d} x\, \sqrt{h} \, K + S_\text{ct}\, ,
\end{split}
\end{equation}	
where $\bdy$ is the asymptotic boundary $\bulk$.
Note in particular that we include a Gibbons-Hawking term on the gluing surfaces $\homsurfA, \homsurfAc$.  If the metric and extrinsic curvature are continuous, these Gibbons-Hawking terms will cancel between the bra and ket parts of the spacetime when computing $S_\text{gr} = S_\text{gr}^k - S_\text{gr}^b$. Continuity of the metric is required by the sewing along $\homsurfA \cup \homsurfAc$, but  continuity of the extrinsic curvature should not be assumed a priori.  For future reference our notation is as follows: $g_{AB}$ denotes the metric in the bulk spacetime $\bulk$, $\gamma_{\mu\nu}$ that on the asymptotic boundary $\bdy$, and $h_{ij}$ the metric induced on the  homology surfaces $\homsurfA \cup \homsurfAc$.  We will later also introduce a metric $q_{IJ}$ on $\fixM$.

Varying the exponent of \eqref{eq:skMn} with respect to the metric at generic points results in the standard Einstein-Hilbert equations of motion. However, we should more carefully consider variations at the various copies of $\Cbulk$ where different branches of the spacetime join.  This is especially true in the vicinity of the splitting surface which, in the limit $n\to 1$, will give us the Hubeny-Rangamani-Takayanagi (HRT) surface \cite{Hubeny:2007xt}. Nevertheless, we begin by addressing the simpler cases of $\homsurfA,\homsurfAc \subset \Cbulk$, saving consideration of the region near $\fixM$ for sections \cref{sec:cover}-\ref{sec:realtime}.

Recall then that our action \eqref{eq:EHact} included Gibbons-Hawking terms at $\homsurfA, \homsurfAc$ for both the bra and ket parts of the spacetime. The gluing conditions at  $\homsurfA, \homsurfAc$  require continuity of the metric, so such terms will cancel between the bra and ket parts if the extrinsic curvature is also continuous.  But, recognizing that the support of the gravitational path integral may include rather wild non-smooth geometries,  we should use the variational principle to derive any conditions on the extrinsic curvature at $\homsurfA, \homsurfAc$.

Having included these Gibbons-Hawking terms, it is straightforward to do so.  As is well-known, working about a spacetime that satisfies the Einstein equations away from $\homsurfA, \homsurfAc$, and $\fixM$, variations that preserve boundary conditions on $\bdy_n$ give
\begin{equation}\label{eq:delSC}
\delta S_\text{gr}[\bulk] = \frac{1}{16\pi G_N} \, \int_{\homsurfA \cup \homsurfAc} \sqrt{h}\,  \pi_{ij} \, \delta h^{ij}\,,  \qquad
\pi_{ij} = K_{ij} - K\, h_{ij}\, ,
\end  {equation}
where $h_{ij}$ is again the induced metric on $\homsurfA \cup \homsurfAc$ and now $\pi_{ij}$ is its the conjugate momentum. Variations of $S_\text{gr} = S_\text{gr}^k-  S_\text{gr}^b$ thus involve the change $\Delta \pi_{ij}$ in $\pi_{ij}$ when passing from a bra to a ket spacetime across $\homsurfA$ or $\homsurfAc$.  Stationarity of $S_\text{gr}$ implies $\Delta \pi_{ij} =0$.  Since the gluing already requires continuity of the induced metric, we see that the extrinsic curvature must be continuous as well.

This fact has important consequences for the causal structure of saddles with replica and CPT symmetry.  As noted at the end of \cref{sec:dmatrep}, imposing both symmetries forces the induced metric to be real on $\homsurfA$ and $\homsurfAc$.  Moreover,
it  requires the extrinsic curvature defined there from the bra side to be the complex conjugate of that defined from the ket side.  But we have seen that stationarity also compels these complex conjugate extrinsic curvatures to agree, so saddle-point geometries must have real extrinsic curvatures on $\homsurfA$ and $\homsurfAc$.  Since the saddle-point geometry will satisfy the usual hyperbolic equations of Einstein-Hilbert gravity away from $\Cbulk$, Cauchy evolution toward the past from $\homsurfA, \homsurfAc$  obliges  the metric to remain real throughout the homology wedges of $\homsurfA, \homsurfAc$ in saddles preserving both replica and CPT symmetries.  This completes the previously advertised argument that symmetric saddles have well-defined homology wedges as expected from more general considerations mentioned above.

As in the Euclidean discussion of \cite{Lewkowycz:2013nqa}, there are two ways to proceed with  specifying the remainder of the variational problem:
\begin{enumerate}[label=\arabic*.,leftmargin=0pt]
\item  We could work with the full $n$-fold geometry $\bulk_n$ which defines  a stationary point of the variational principle satisfying the full asymptotic boundary conditions $\bdy_n$.  For reasons that will be clear below, we henceforce refer to $\bulk_n$ as the \emph{covering space} geometry.  If desired, we may attempt to simplify the problem of solving the equations of motion by imposing on $\bulk_n$ any of the symmetries described above.
\item Alternately, if we are interested only in symmetric saddles we can impose ${\mathbb Z}_n$ replica symmetry from the outset and consider the quotient spacetime $\widehat{\bulk}_n = \bulk_n/\mathbb{Z}_n$.  Clearly, $\bulk_n$ is an $n$-fold cover of $\widehat{\bulk}_n$. We will refer to $\widehat{\bulk}_n$ as a single \emph{fundamental domain} below following the terminology of \cite{Haehl:2014zoa}. The fundamental domain will include a single bra branch and a single ket branch.  Note that there is only one splitting surface $\fixM$ in $\bulk_n$, so it must remain fixed under the action of the replica symmetry group.  As a result, one expects the quotient spacetime to  have an explicit source of curvature (a cosmic brane) localized along $\fixM$. As in \cite{Lewkowycz:2013nqa}, the fundamental domain  description is particularly useful in analytically continuing to non-integer $n$ (when effects that break replica symmetry can be ignored).
\end{enumerate}

Note that the covering space perspective is the more general of the two, in that it also allows discussion of saddles that break replica symmetry.  Such saddles have recently been shown to be important near phase transitions \cite{Penington:2019kki,Dong:2020iod,Akers:2020pmf}. Furthermore, working in the covering space provides the cleaner description of the physics as it avoids introducing artificial singularities at $\fixM$.  We therefore focus on this perspective through \cref{sec:realtime} below.  However, we will return to the quotient $\widehat{\bulk}_n = \bulk_n/\mathbb{Z}_n$ in \cref{sec:fundomain} since (as in \cite{Lewkowycz:2013nqa}) in simple cases this perspective greatly simplifies analytic continuation to non-integer $n$.

\subsection{Contributions to the action from the splitting surface}
\label{sec:cover}

It remains to discuss contributions to $S$ in \eqref{eq:TotalS} from the region near the cosmic brane $\fixM$.  We focus here on defining such contributions, postponing to sections \cref{sec:imtime} and \cref{sec:realtime} the description of boundary conditions at $\fixM$ that make the variational principle well-defined.   In particular, here we wish to understand whether there are delta-function-like contributions to the Ricci scalar at $\fixM$ that should be taken to give finite contributions $S_{\fixM}$ to $S_\text{gr}$.\footnote{We should emphasize here that our preceeding discussion makes clear that there are no singularities along the past light-cone of $\fixM$. The delta function curvature singularities are localized solely on the codimension-$2$ fixed point locus.}  At first sight, the situation may appear especially confusing due to the fact that $\fixM$ lies at the boundary between the bra and ket parts of the spacetime, and thus at the locus where the action changes sign.

However, things are simpler than they appear. Since we expect no strong curvatures from the directions along the brane, any such delta-functions contributions will be associated with the metric transverse to the brane. The problem thus becomes effectively two-dimensional in the plane normal to $\fixM$, which we recall has Lorentz signature.  We may thus  {\it define} the contribution $S_{\fixM}$ from a small tubular region $\mathscr{U}_\epsilon$ containing $\fixM$ by using a generalization of the Gauss-Bonnet theorem.
Topologically, we require $\mathscr{U}_\epsilon$ to be the product ${\mathscr{\tilde U}}_\epsilon \times \fixM$ for an appropriate two-dimensional space ${ \mathscr{\tilde U}}_\epsilon$, so that we may apply our Gauss-Bonnet theorem to the latter.

The idea of using a generalization of the Gauss-Bonnet theorem was suggested in e.g., \cite{Louko:1995jw}.  However, we use a slightly different generalization due to the fact that $\fixM$ lies on a timefold. The required generalization may be motivated by analytically continuing a Euclidean metric as $t_{_\text{E}} \rightarrow it_{_\text{L}}$ in the ket parts of the spacetime, but using the complex conjugate $t_{_\text{E}} \rightarrow -it_{_\text{L}}$ in the bra parts of the spacetime.  This is the same change of sign that creates the timefold as it gives $t_{_\text{L}} < 0$ in both the bra and ket parts (though we think of the bra parts as coming from $t_{_\text{E}} >0$).  It is then associated with particular signs in our Gauss-Bonnet theorem.  As in \cite{Louko:1995jw}, the resulting Gauss-Bonnet theorem will also involve various factors of $i=\sqrt{-1}$.

To describe the relevant signs, it is useful to introduce $\eta=\pm 1$ taking the positive sign on the ket parts of the spacetime and taking the negative sign on the bra parts of the spacetime. This allows us to define the action contribution from $\mathscr{U}_\epsilon$ by
\begin{equation} \label{eq:cGB}
\begin{split}
iS_{\fixM}
&: =
	 \lim_{\epsilon \rightarrow 0} \frac{i}{16\pi G_N} \int_{\mathscr{U}_\epsilon} \,  d^{d+1}x\  \eta\,\sqrt{-g} R \\
&:=
	\lim_{\epsilon \rightarrow 0}  \frac{-i}{8\pi G_N}  \,\int_{\partial \mathscr{U}_\epsilon} \, d^dx\ \eta\,\sqrt{|h|}\, K
		+  \frac{1}{4G_N} \, \chi({\mathscr{\tilde U}}_\epsilon) \, A_\fixM.
\end{split}
\end{equation}

The extrinsic curvature term in \eqref{eq:cGB} is subtle and requires detailed comment.  First, in Lorentz signature it receives imaginary delta-function-like contributions from submanifolds of $\partial \mathscr{U}_\epsilon$ where $\partial \mathscr{U}_\epsilon$ changes from spacetime to timelike (or vice versa). This is most easily seen in the $1+1$ case where $K$ can be written as the derivative of the boost parameter $\theta$ (aka rapidity) describing the tangent to $\partial \mathscr{U}_\epsilon$.  Since null tangents have $\theta = \infty$, the boost parameter $\theta$ has a pole at such transitions.  Furthermore, in the ket part of the spacetime and when a timelike portion of $\partial \mathscr{U}_\epsilon$ is attached to a spacelike portion of $\partial \mathscr{U}_\epsilon$ that lies to its past, the above analytic continuation recipe makes $\theta$ real  on the timelike portions (where the normal is spacelike) but yields $\theta \in i \frac{\pi}{2} + {\mathbb R}$ on spacelike portions (where the normal is timelike).  Integrating through the transition in a ket part of the spacetime thus yields a finite contribution $i\frac{\pi}{2}$, or $-i\frac{\pi}{2}$ in the alternate case where the timelike portion of $\partial \mathscr{U}_\epsilon$ is attached to a spacelike portion of $\partial \mathscr{U}_\epsilon$ that lies to its future.  In higher dimensions the contribution is similarly $\pm i\frac{\pi}{2} A_{\parallel}$, where the longitudinal area $A_{\parallel}$ becomes $A_{\fixM}$ in the limit $\epsilon \rightarrow 0$.  Contributions from the bra parts of the spacetime take the complex-conjugate form.   These localized contributions (from both the ket and bra parts) are shown explicitly as the last term in \eqref{eq:cGB} and are therefore no longer included in the extrinsic curvature term there.

\begin{figure}[h!]
\begin{subfigure}[a]{\textwidth}
\centering
\includegraphics[scale=0.4]{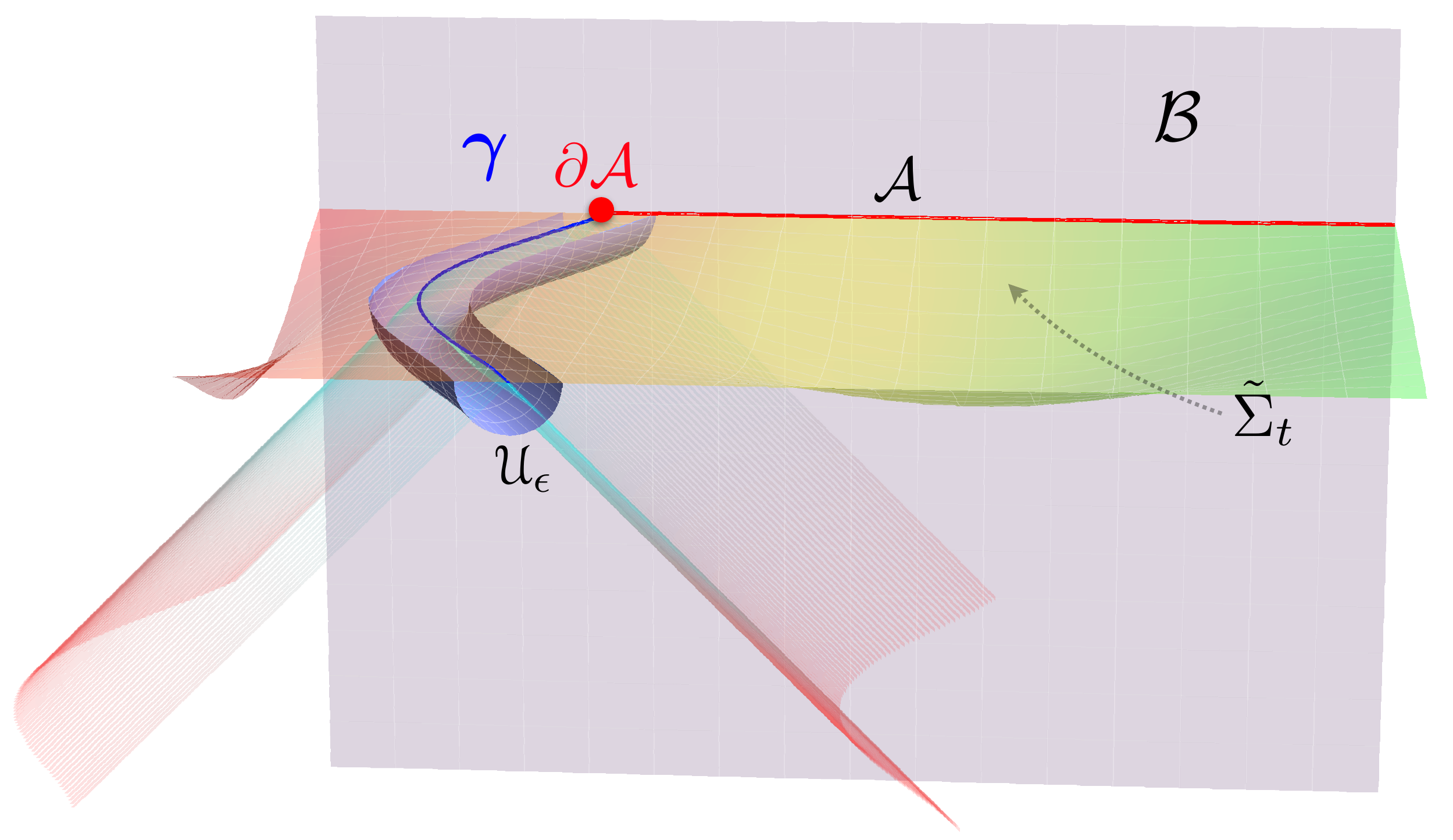}
\caption{Regulating the fixed point locus in the spacetime by excising a tubular neighbourhood around it. \cref{fig:epsregulator} displays a slice through this at fixed $y^I$ for ease of visualization.}
\label{fig:fixbc}
\end{subfigure}
\begin{subfigure}[b]{\textwidth}
\centering
\begin{tikzpicture}[scale=1.3]
\foreach \x in {0}
{
\draw[red,thick]  (\x+1,0) .. controls (\x+2,-0.3) and (\x+3,+0.5) .. (\x+4,0);
\draw[brown,thick] (\x-1,0) .. controls (\x-2.5,+0.2)  and (\x-3,-0.5) .. (\x-4,-0.25);
\draw[red,thin, dashed]  (\x,0) .. controls (\x+0.5,0.2) and (\x+0.9,+0.1) .. (\x+1,0);
 \draw[brown,thin,dashed] (\x,0) .. controls (\x-0.5,-0.2)  and (\x-0.9,-0.1) .. (\x-1,0);
\draw[black] (\x-2,-2) -- (\x,0) -- (\x+2,-2);
\draw[blue, fill =blue] (\x,0) circle [radius = 3pt];
\foreach \xend in {\x+1, \x-1}
{
	\draw[black,fill=black] (\xend,0) circle [radius = 1pt];
}
\draw[thick, blue!40] (\x-1,0) arc (-180:0:1cm);
\node at (\x+2,0) [above=2pt]{$\homsurfAreg$};
\node at (\x-2,0) [above]{$\homsurfAcreg$};
\node at (\x,0) [above] {$\fixM$};
\node at (\x,-0.4){$\mathscr{U}_\epsilon$};
\node at (\x+1.1,-0.45) [below]{$\color{blue}{\partial\tilde{\mathscr{U}}_\epsilon}$};
\node at (\x,-1.2) [below]{$\Cbulk^\epsilon$};
\draw[dotted,black, ->] (\x-2.5,-0.1) ..  controls (\x-2, -1) and (\x-1.3, -1.2) .. (\x-0.2,-1.5);
\draw[dotted,black, ->] (\x+2.7,0.1) ..  controls (\x+2.3, -1) and (\x+1.8, -1.2) .. (\x+0.2,-1.5);
}
\end{tikzpicture}
\caption{The local geometry in the cosmic brane excised spacetime shown at fixed $y^I$. The inner boundary consists of half of
$\partial \tilde{\mathscr{U}}_\epsilon$ (blue), along with two regions we call $\homsurfAreg, \homsurfAcreg$ (solid red and brown lines). The latter are regulated versions of $\homsurfA, \homsurfAc$ with the differences indicated by dashed lines.   The Gibbons-Hawking term associated with $\homsurfA \cup \homsurfAc$ in \eqref{eq:EHact} should understood as the $\epsilon \rightarrow 0$ limit of the integral over $\homsurfAreg \cup \homsurfAcreg$.}
\label{fig:epsregulator}
\end{subfigure}
\caption{The local geometry near the cosmic brane and the boundary conditions in Lorentz signature.  We have exhibited here the regulated ket spacetime $\bulkket_\epsilon$ which is obtained from $\bulkket$ by excising a topological half-disc ribbon around the fixed point locus $\fixM$ (i.e., half of the exicison domain $\mathscr{U}_\epsilon$).}
\label{fig:regfixeps}
\end{figure}
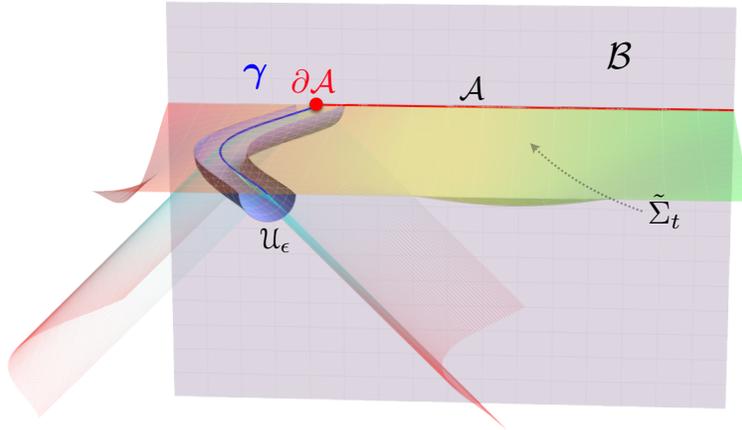
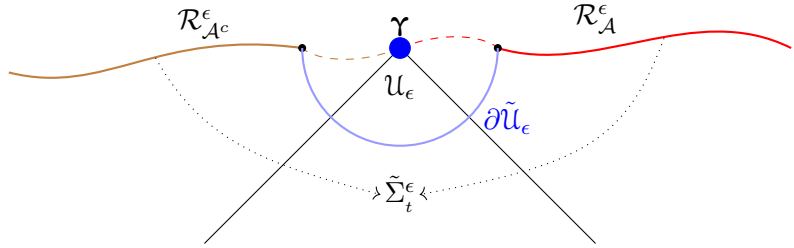

The other subtlety involves the possibility of explicit corner contributions to the extrinsic curvature term when $\partial \mathscr{U}_\epsilon$ is not smooth.  In particular, such terms arise if $\mathscr{U}_\epsilon$ fails to have orthogonal intersection with $\homsurfA \cup \homsurfAc$ so that there is an abrupt change in the normal to $\partial \mathscr{\tilde U}_\epsilon$ when crossing from a ket spacetime to the attached bra spacetime.\footnote{This can  happen with  the surface $\partial \mathscr{\tilde U}_\epsilon$ either remaining entirely  spacelike, or having both spacelike and timelike pieces. In  the latter case, there  are the contributions from the change in the signature  described above. In addition one also encounters corner terms from the intersection of  the timelike part of $\partial \mathscr{\tilde U}_\epsilon$ with $\homsurfA \cup \homsurfAc$. These however give a real contribution and thus cancel out in between the ket  and the  bra.\label{fn:generalcorner}}
 The interested reader can find a discussion of such terms in  \cref{app:alt}, but for the moment we simply choose $\partial \mathscr{U}_\epsilon$ smooth to avoid such terms.\footnote{A good general discussion can be found in \cite{Jubb:2016qzt} though such contributions have  been  considered various earlier discussions eg., \cite{Farhi:1989yr,Neiman:2013ap,Dong:2016hjy}.}  In particular, we take $\partial \mathscr{U}_\epsilon$ to be of the form depicted by the blue surface in  \cref{fig:fixbc}.

The extrinsic curvature thus gives both real and imaginary contributions to \eqref{eq:cGB}.  However,  for spacetimes that are symmetric under both the ${\mathbb Z}_n$ cyclic symmetry and the conjugation symmetry, the imaginary contributions from the extrinsic curvature term cancel between the bra and ket parts of the spacetime.  Such spacetimes thus have real weight $e^{iS}$ as required by the conjugation symmetry.   As discussed in \cref{app:thermal}, the spacetime of \cite{Jana:2020vyx} realizes the framework described here in an $n=1$ Schwinger-Keldysh example that illustrates the reality of $e^{iS}$  in a particularly explicit manner.

Before proceeding, we emphasize that  \eqref{eq:cGB} is to define the action in the full region $\mathscr{\tilde U}_\epsilon$.
Thus while the Gibbons-Hawking term in \eqref{eq:EHact} was written as an integral over $\homsurfA \cup \homsurfAc$, for consistency it should be understood as being defined by integrating only over the parts $\homsurfAreg \cup \homsurfAcreg$ of $\homsurfA \cup \homsurfAc$ that lie outside ${\mathscr{U}}_\epsilon$ and then taking the limit $\epsilon \rightarrow 0$.

Due to the above subtleties, in calculating the path integral weight for a  conjugation-symmetric saddle it will often be simplest to compute just the ket contributions to $S$ (including the ket contributions to \eqref{eq:cGB}), and then to use the conjugation symmetry to write the bra contributions as the complex conjugates.  In doing so, one may wish to employ an alternate accounting scheme in which the splitting surface contribution $\eqref{eq:cGB}$ is absorbed into the contributions from the ket and bra parts of the spacetime.  This alternate scheme is described in \cref{app:alt}.  We also discuss there the inclusion of explicit corner terms that necessarily arise if one wishes to avoid the presence of timelike pieces of $\partial \mathscr{U}_\epsilon$.

It now remains to evaluate \eqref{eq:cGB} on interesting configurations and to understand the implications for our variational principle.  Our approach will be to use previous work on the Euclidean replica problem to motivate a useful ansatz for the metric near $\fixM$, and then to check that this ansatz allows for stationary points of the full action $S$ under variations of the metric near $\fixM$.

\subsection{Imaginary-time boundary conditions at the splitting surface}
\label{sec:imtime}

As stated above, we will first review the boundary conditions at $\fixM$ associated with the analogous variational problem in Euclidean signature.  Although we are interested here in the covering space perspective, and while the covering space is smooth in Euclidean signature, it will nevertheless be useful to allow a localized delta-function source of curvature at $\fixM$ in the (generally off-shell) metric configurations that we consider.  As is well known, this gives a conical defect, though the condition that the defect should vanish on-shell can be recovered from an equation of motion associated with variations with respect to the area $A_{\fixM}$ of $\fixM$.  This perspective will be useful in the real-time context where the real Lorentz-signature replica wormholes require causal singularities so that, with replica boundary conditions, there are simply no smooth Lorentz-signature metrics on which to base a discussion of the real-time variational problem.  Interpreting the vanishing-defect condition as an equation of motion rather than a boundary condition will thus allow us to formulate a variational principle on the original contour of integration defined by real Lorentz-signature metrics, even if the final stationary points can be accessed only by deforming this contour into the complex plane.

We parameterize the Euclidean conical defect by using  $2\pi m$ to denote the full angle around the cone, by which we mean that small circles of radius $r$ about the defect have circumference $2\pi m r$.  Thus the case $m=1$ is smooth.  For later use, we allow $m$ to be independent of the replica number $n$.

As discussed in \cite{Unruh:1989hy}, a convenient set of quasi-cylindrical coordinates generalizing the notion of Gaussian or Riemann normal coordinates can be constructed using geodesics launched orthogonally from the conical defect.  We pick $(r,t_{_\text{E}})$ to parameterize the normal plane and $y^I$ to be longitudinal along the cosmic brane. In such coordinates, with the cosmic brane located at $r=0$,  the metric can be taken to have the form
\begin{equation}
\label{eq:met}
ds^2 = dr^2 + \left(m^2+\order{r^\alpha}\right) r^2 \, dt_{_\text{E}}^2 + \order{r^0}\,  dy^I dy^J + \order{r^2}\, dt_{_\text{E}} dy^I\, .
\end{equation}
Here $t_{_\text{E}}$ is an angular coordinate taking values in $[0,2 \pi)$ and the $y^I$ denote an arbitrary set of coordinates on the conical defect.
We take the rates  of radial fall-off to be as prescribed in \eqref{eq:met} with the constraint $\alpha > 1$.  The metric coefficients may contain arbitrary functions of
$(r,t_{_\text{E}},y^I)$ subject to periodicity under $t_{_\text{E}} \to t_{_\text{E}} + 2\pi$. As a result, all the functions can be expanded in a Fourier series involving only integral powers of $e^{i t_{_\text{E}}}$.

It will be useful below to pass to complex coordinates.  If this is an $n$-replica calculation, it will be useful to choose coordinates $v, \bar v$ adapted to a single replica in the sense that $v, \bar v$ are periodic in $t_{_\text{E}}$ with period $\frac{2\pi}{n}$, and also such that $v, \bar v$ are real on the CPT-invariant surfaces.\footnote{Thus our $v,\bar{v}$ are not the $z,\bar z$ of \cite{Dong:2019piw}.}  In particular, we take  $v=r^{\frac{1}{\hat m}}\, e^{int_{_\text{E}}}$ with $\hat{m} := \frac{m}{n}$.  In terms of this $v, \bar v$, the metric \eqref{eq:met} becomes  
\begin{equation}\label{eq:mv}
\begin{split}
ds^2 &= \hat{m}^2 (v \bar v)^{\hat{m}-1}\, dvd\bar{v} + T \,\frac{(\bar{v} \,dv-v\, d\bar{v})^2}{(v\bar{v})^{2-\hat m}}
+ q_{IJ} \, dy^I dy^J + 2i\, W_J \, dy^J \, \frac{\bar{v} dv-v d\bar{v}}{v \bar{v}},\\
T&=\order{r^\alpha}  = \order{(v\bar{v})^{\frac{\alpha m}{2n}}} =  \order{(v\bar{v})^{\frac{\alpha \hat{m}}{2}}} \,,\quad
q_{IJ}=\order{r^0}=\order{(v \bar{v})^0 }\,,\\
W_J&= \order{r^2} = \order{(v \bar{v})^{\frac{m}{n}}} = \order{(v \bar{v})^{\hat{m}}}\, .
\end{split}
\end{equation}
Since $v \bar{v}$ is real and positive, we may define fractional powers (e.g., $(v \bar{v})^{\hat m}$) to be real and positive as well.  Note that periodicity in $t_{_\text{E}}$
requires the metric functions $T,q_{IJ},W_J$ to contain only integer powers of $v^{1/n}$ and $\bar{v}^{1/n}$, except where they appear in the manifestly real combination $v\bar{v}$.  Furthermore, replica symmetry solutions will involve only integer powers of $v, \bar v$ (except in the combination $v\bar v$).

Having specified the local geometry \eqref{eq:mv} in the vicinity of the splitting surface, we can proceed to show that Einstein-Hilbert variational problem is well-defined. To this end, let us first focus on the modified variational problem defined as in \cite{Lewkowycz:2013nqa} by using the Einstein-Hilbert action but removing the delta-function curvature contribution normally  associated with the conical singularity. We denote the resulting brane-excised Euclidean action by $\check{I}_{m,n}$.  One may think of $\check{I}_{m,n}$ as being defined by simply dropping the $S_{\fixM}$ term from \eqref{eq:TotalS} without changing the definition of the other terms.\footnote{In particular, at finite regulator $\epsilon$ there will still be no Gibbons-Hawking term at $\partial \mathscr{U}_\epsilon$.}    Although the parameters $m, n$ play very different roles -- with $m$ controlling the conical singularity and $n$ controlling only our description of the spacetime through the definition of the coordinates $v, \bar v$ -- it will be useful to keep both labels in $\check{I}_{m,n}$ for comparison with the real-time discussion below.

The analysis of \cite{Dong:2019piw} then shows that, for fixed $m$ and $n$, requiring the behavior \eqref{eq:mv} suffices to make $\check{I}_{m,n}$ a good variational principle for the Einstein equations.\footnote{\label{foot:ext} This was understood much earlier in many contexts (see e.g. \cite{Geroch:1987qn,Lewkowycz:2013nqa}), though we are not aware of a fully general prior study of the case $m>1$. We also comment that this variational principle provides what one may call an equation of motion for the splitting surface given by equation (A.66) of \cite{Dong:2019piw}, that in some sense determines the location of $\fixM$ relative to other features of the spacetime.  For example, in the limit $n\rightarrow 1$, this condition requires $\fixM$ to be extremal.}   Indeed, although different coordinates were used to describe the covering spacetime, our $\check{I}_{m,n}$ agrees precisely with the object called $\tilde{I}_{m}$ in \cite{Dong:2019piw}.
We also recall from their analysis that  the Hamilton-Jacobi variation of the brane-excised Einstein-Hilbert action with respect to $m$ is just $-\frac{A_{\fixM}}{4G_N}$, where $A_{\fixM}$ is the area of the cosmic brane $\fixM$
\begin{equation}
\label{eq:checkmvary}
-\fdv{\check{I}_{m,n}}{m} = \frac{A_{\fixM}}{4G_N} .
\end{equation}	
In particular, this holds for arbitrary real $m$.

As a result, if one happened to be interested in a related problem that fixed the area $A_{\fixM}$ of $\fixM$ but left the opening angle $2\pi m$ of the conical singularity unconstrained, one could construct a good variational principle $I_A$ for that problem via the Legendre transform
\begin{equation}
\label{eq:fixAE}
{I}_A = \check{I}_{m,n} + (m-1)\,\frac{A_{\fixM}}{4G_N}\,.
\end{equation}
The ambiguity in the Legendre transform (associated with adding an arbitrary $m$-independent function of $A_{\fixM}$) has been fixed by requiring the value of \eqref{eq:fixAE} to agree with that of the standard Einstein-Hilbert action in the case $m=1$ (when there is no conical singularity).  This then has the consequence that \eqref{eq:fixAE} in fact agrees with the standard Einstein-Hilbert action for all $m$.  In particular, for any $m$  the term $(m-1)\,A_{\fixM}$ in \eqref{eq:fixAE} precisely restores the contribution $S_{\fixM}$ associated with a possible conical defect.

The above argument provides an especially clean way to see that, if one uses the standard Einstein-Hilbert action, requiring stationarity under constrained variations that fix $A_{\fixM}$ imposes the equations of motion away from $\fixM$ but allows a general conical singularity at $\fixM$. If one further also requires stationarity under variations of $A_{\fixM}$ (which then amount to Hamilton-Jacobi variations of \eqref{eq:fixAE} with respect to $A_{\fixM}$), one obtains the condition $m=1$ that requires the spacetime to be smooth.

In summary, we see that an alternate way to impose the condition that the full $n$-replica geometry be smooth at $\fixM$ is simply to require that it provide a stationary point of the Einstein-Hilbert action \eqref{eq:fixAE}  with respect to variations of $A_{\fixM}$.  In particular, from \eqref{eq:checkmvary} we see that the action \eqref{eq:fixAE} is also stationary under variations of $m$ so that we may consistently treat the smoothness condition $m=1$ as an equation of motion that follows from the gravitational dynamics rather than as an a priori boundary condition.  Indeed, since the support of any path integral measure will not be concentrated on smooth configurations, it is natural to take the equation-of-motion perspective to be more fundamental than simply imposing smoothness as a boundary condition.

\subsection{Real-time boundary conditions at the splitting surface}
\label{sec:realtime}
With this understanding we can now turn to the real-time context.   While we wish to avoid analytic continuation of our physical metric, its boundary conditions, or any relevant real-time sources, we are nevertheless free to use the Euclidean analysis as {\it motivation} for a choice of real-time boundary conditions  at $\fixM$.  As in the imaginary time case, we will first discuss a brane-excised variational principle which in some sense allows an arbitrary fixed `defect' at $\fixM$.  We will then use this brane-excised action to show that similar boundary conditions promote the $S$ of \eqref{eq:TotalS} to a well-defined variational principle that can be formulated on real Lorentz-signature replica wormhole spacetimes.  In this latter variational principle, the defect parameter $m$ is free to vary, but is then determined on-shell by an equation of motion associated with varying the area $A_{\fixM}$ of the splitting surface.  Furthermore, while real replica wormholes cannot make the action stationary at $\fixM$, we will see that complex replica wormholes can do so, and that they are also compatible with the full set of stationarity conditions at the level of counting the relevant equations.  This sets the stage for the construction of examples in the companion paper \cite{Colin-Ellerin:2020exa}, which will demonstrate that  the desired complex saddles do in fact exist (at least in the contexts studied there).

To begin our real-time analysis, we remind the reader that we consider here the real-time covering-space description of the $n$-replica saddle, which thus has $n$ bra spacetime pieces and $n$ ket spacetime pieces glued together along appropriate surfaces $\homsurfA, \homsurfAc$.  For the moment, we do not require any particular symmetries of this spacetime.

Let us first focus on one of the ket spacetimes (which is associated with the more familiar $e^{iS^k_\text{gr}}$ in the path integral), with the understanding that the complex-conjugate boundary conditions will hold on the bra spacetimes.
As stated above, our Lorentzian approach will be motivated by Wick-rotation of the Euclidean results discussed in \cref{sec:imtime}.  In particular, it is natural to associate any particular pair of bra and ket spacetimes with a single fundamental domain of some replica-symmetric Euclidean solution under the ${\mathbb Z}_n$ cyclic symmetry.  Recall that the Euclidean metric near $\fixM$ in any such fundamental domain can be described by \eqref{eq:mv}.  Now, note that taking the above Euclidean complex coordinates $v, \bar{v}$ to be $x+i \,t_e, x-i \,t_e$ and Wick-rotating $t_e \to i t$ would define
`light-cone' coordinates\footnote{ This choice differs by signs from the standard null coordinates $x^\pm \equiv t \pm x = \pm \tx^\pm$. It is also worth noting that the Euclidean time coordinate $t_e$ here is a Cartesian coordinate, in contrast to the angular coordinate $t_{_\text{E}}$ used in \eqref{eq:met}.}
$\tx^\pm = x\pm t$.  Applying this transformation $v\to \tx^- $ and $\bar{v} \to \tx^+$   to \eqref{eq:mv} yields the metric (with $\hat{m} \equiv \frac{m}{n}$)
\begin{equation}\label{eq:mxL}
\begin{split}
ds^2 &=
	\sigma(\tx^+,\tx^-)\, d\tx^ + d\tx^- + T \,\frac{(\tx^+ \,d\tx^--\tx^-\, d\tx^+)^2}{(\tx^+\tx^-)^{2-\hat m }} \\
&\qquad \qquad 	
	+ q_{IJ} \, dy^I dy^J + 2\, W_J \, dy^J \, \frac{\tx^+ d\tx^--\tx^- d\tx^+}{\tx^+ \tx^-},\\
\text{with} & \\
\sigma&(\tx^+, \tx^-)
\equiv
	\hat{m}^2 (\tx^+ \tx^-)^{\hat{m}-1}\, ,\\
T&=
	\order{r^\alpha}  = \order{(\tx^+\tx^-)^{\frac{\alpha m}{2n}}} =  \order{(\tx^+\tx^-)^{\frac{\alpha \hat{m}}{2}}} \,,\\
q_{IJ}
&=
	\order{r^0}=\order{(\tx^+\tx^-)^0 }\,,\\
W_J&=
	\order{r^2} = \order{(\tx^+\tx^-)^{\frac{m}{n}}} = \order{(\tx^+\tx^-)^{\hat{m}}}\,.
\end{split}
\end{equation}

Furthermore, the analogue of the Euclidean metric being periodic in $t_{_\text{E}}$ with period $2\pi$ is to require  that the metric functions $T,\,q_{IJ} ,\, W_J$  involve only integer powers of $(\tx^\pm)^{\frac{1}{n}}$, except perhaps in the combination $\tx^+ \tx^-$.  In other words, we require these coefficients to be functions of the triple $((\tx^+)^{\frac{1}{n}},(\tx^-)^{\frac{1}{n}},\tx^+ \tx^-)$ that are analytic in the first two arguments in some neighborhood of the origin $\tx^+ = 0 = \tx^-$.  Note that such local analyticity is to be expected at any source-free regular point of the equations of motion and does not restrict the use of non-analytic sources at the asymptotic boundary.  (Replica-symmetric solutions will involve only integer powers of $\tx^\pm$, again with the possible exception of appearance in the combination $\tx^+ \tx^-$.)

Let us first consider the case where $\hat m$ is a positive integer, and in particular the case $\hat m=1$ (where $m=n$).  For appropriate $T$, $q_{IJ}$, and $W_J$, the metric \eqref{eq:mxL} can then be both real and completely smooth away from the timefold.  In particular, this case includes real Lorentz-signature replica wormhole spacetimes of the sort that define the domain of integration for the Lorentz-signature path integral described in \cref{sec:dmatrep}.

On the other hand, for more general values of $\hat m$ (or for more general choices of  $T$, $q_{IJ}$, and $W_J$), metrics satisfying \eqref{eq:mxL} are at best a complex deformation of the above real Lorentz-signature replica wormholes.  This deformation will be of interest below, though it has several subtleties that require comment.
\begin{enumerate}[label=(\alph*)., leftmargin= 1cm]
\item The first subtlety is that \eqref{eq:mxL} can involve negative powers of $\tx^+ \tx^-$, in which case it is singular when either $\tx^+$ or  $\tx^-$ vanish.  More generally, we should expect a solution to the Einstein equations that behaves like \eqref{eq:mxL} near $\fixM$ to be singular on the past light cone of $\fixM$.    This is an interesting difference from the Euclidean case, where the metric is smooth away from the tip of the cone.  Note that for $m=1$ (and thus for $\hat{m} = \frac{1}{n}$) the singularity was obtained by making a complex coordinate transformation and analytically continuing a smooth Euclidean solution, so in that case this appears to be a form of coordinate singularity.  It should thus be harmless for $m=1$, though see further discussion in \cref{sec:Disc}.  Note that for $n> 1$ the choice $m=1$ forbids taking $\hat m= \frac{m}{n}$ to be a positive integer, and is thus intrinsically complex.
\item The second subtlety is that, since $\tx^+$ and $\tx^-$ can be negative (and in particular since $\tx^+\tx^-$ is negative in the Milne wedge), fractional powers can be complex and require appropriate definition.
\end{enumerate}

In fact, motivated by the Euclidean description, we will deal with both issues in much the same way, choosing our ket spacetime definitions for general $\hat m$, $T$, $q_{IJ}$, $W_J$ to match what would be obtained by analytically continuing $t$ through the upper half-plane (since $t_E \rightarrow it$ and the ket part of the spacetime comes from $t_E<0$.).  This amounts to introducing the $i\varepsilon$ prescriptions $\tx^\pm \rightarrow \tx^\pm \mp i \varepsilon$ (with $\varepsilon >0$) and taking the powers of $\tx^\pm$ appearing in \eqref{eq:mxL} to be analytic functions. Thus for negative $\tx^+$ we have
$(\tx^+)^\frac{1}{n} = e^{-\frac{i\pi}{n}} \,  \abs{\tx^+}^\frac{1}{n}$
and for negative $\tx^-$ we find
$(\tx^-)^\frac{1}{n} = e^{+\frac{i\pi}{n}} \,\abs{\tx^-}^\frac{1}{n}$.
In particular, $(\tx^+ \tx^-)^{\frac{1}{n}}$ remains real at positive at $t=0$, so that \eqref{eq:mxL} is compatible with the previously advertised requirement that the metric and its extrinsic curvature be real and positive on $\homsurfA$ and $\homsurfAc$.  Nevertheless, it is forced to be complex in the Milne wedge to the `past' of $\fixM$.

We take the above specifications to be part of the boundary conditions at $\fixM$ for our real-time variational problem.  In particular, this refines the definition of the real Lorentz-signature replica wormhole configurations that specify the original domain of integration for our Lorentz-signature path integral.  As noted above, reality generally requires $\hat m$ to be an integer.  Metrics with other values of $\hat m$ simply do not lie in the original domain of integration.

Nevertheless, metrics with general $\hat m$ are allowed to appear in complex deformations of that domain.  In particular,
the $i\varepsilon$ prescription included in our proposal for general $\hat m$ turns out to be very useful.  For any fixed $m,n$  it will imply that the above  are indeed a valid set of boundary conditions for a variational problem associated with the brane-excised Einstein-Hilbert action $\check{S}_{m,n} := S - S_{\fixM}$ on the timefolded bulk spacetime. This timefolded spacetime only retains the  $t\le0$ regions in both the ket and bra spacetimes, and we may think of the excision as removing a small disk-shaped region of size\footnote{Since we are Lorentz signature, this `size' does not in any way measure proper distance from $\fixM$.} $\epsilon$ around the origin $\tx^\pm =0$, though we eventually take $\epsilon \to 0$ at the end of the computation.\footnote{ The reader might find it helpful to recall our definition $I = -iS$ below \eqref{eq:renyiMn}, which makes it natural to talk about $I$ for the Euclidean computation, but stick to $S$ for the Lorentzian on-shell action.  }

The fact that $\check{S}_{m,n}$ gives a valid variational principle for fixed $m,n$ can be read directly from the Euclidean analysis of \cite{Dong:2019piw}.  Since the arguments of that reference simply manipulated power series expansions in their $z^\frac{1}{n}, (\bar z)^\frac{1}{n}$, it is immediate that every step of can be `Wick rotated' and rewritten in terms of $\tx^\pm$ to yield an equally-valid treatment of the real-time boundary conditions above. And as in \cref{foot:ext}, this in a certain sense provides an equation of motion for the splitting surface $\fixM$.  Furthermore, by the same argument we can read from \cite{Dong:2019piw} that the Hamiltonian-Jacobi variation of $\check{S}_{n,m}$ with respect to $m$ yields
\begin{equation}
\label{eq:LHJm}
\fdv{\check{S}_{n,m}}{m} = -i \frac{A_{\fixM}}{4G_N} \, \, .
\end{equation}	

Now, we in fact wish to show that metrics of the form \eqref{eq:mxL} promote the full `Einstein-Hilbert action' $S = \check{S}_{m,n} + S_{\fixM}$ to a well-defined variational principle.  And as in the imaginary-time context, we will now allow $m$ to vary as well.   We thus need to consider the piece $S_{\fixM}$ associated with the region near $\fixM$.  To this end, in some ket part of our spacetime, let us consider a small semicircle $\mathscr{C}_\epsilon$  about the origin
in the lower half of the (real) $\tx^\pm$ plane (i.e., of the form described by the orange curve in \cref{fig:epsregulator}). In particular, we take the curve to have orthogonal intersection with $\homsurfA, \homsurfAc$ so that there is no delta-function contribution to the right-hand-side of \eqref{eq:cGB} from `corners' where the normal to $\mathscr{C}_\epsilon$ changes when crossing from a ket to a bra spacetime across $\homsurfA, \homsurfAc$.

In the limit of small $\epsilon$, non-vanishing contributions to the integrated extrinsic curvature will come only from the explicit terms in \eqref{eq:mxL}; terms associated with $T$, $q_{IJ}$, and $W_J$ decay too quickly to contribute.  By inspection, one sees that the explicit terms in \eqref{eq:mxL} are invariant under both replica and conjugation symmetry.  This means that their contributions to the real part of the integrated extrinsic curvature must cancel between the bra and ket parts of our spacetimes.  We may thus focus on the imaginary parts.

Now, as noted below \eqref{eq:cGB}, there are in fact two kinds of contributions to $\Im  \int_{ \mathscr{C}_\epsilon} \, d^dx\, \sqrt{h}\, K$,
which is to be computed using the outward pointing unit normal. The first comes from computing the extrinsic curvature in the past Milne region where the metric can be complex.  But the second is associated with loci where the spacetime metric is real and of Lorentz signature but $\mathscr{C}_\epsilon$ transitions from being spacelike to being timelike (or vice versa).  Noting that the relevant terms in \eqref{eq:mxL} are real outside the past Milne region, we choose $\mathscr{C}_\epsilon$  to coincide with the surface $\tx^- - \tx^+ = \epsilon$ in the past Milne region (and a bit outside) but to otherwise be an arbitrary smooth curve that meets the timefold orthogonally.  This in particular requires $\mathscr{C}_\epsilon$  to be spacelike just outside the past Milne wedge (and on either side) but timelike where it meets the timefold. Since we currently focus on a ket part of the spacetime, the contribution to $\Im  \int_{ \mathscr{C}_\epsilon} \, d^dx\,\sqrt{h}\, K$ from outside the past Milne wedge is thus
\begin{equation}\label{eq:ImK1}
-\Im  \int_{ \mathscr{C}_\epsilon} \, d^dx\,\sqrt{h}\, K \bigg|_{\text{outside past Milne wedge}} =
	2 \times \frac{\pi}{2} \times A_{\fixM} = \pi A_{\fixM}
\end{equation}	
 from the associated transitions.

The contribution to $\Im \int_{ \mathscr{C}_\epsilon} \, d^dx\, \sqrt{h}\, K$ from the past Milne wedge is also straightforward to compute.  Since $\sqrt{h}K$ is real for any spacelike surface outside the Milne wedge (again restricting to contributions from the explicit terms in \eqref{eq:mxL} both here and below), we may in fact integrate $\sqrt{h}\, K$ over any surface cutting across the Milne wedge that becomes spacelike in the Rindler wedges.  For simplicity, we choose this to be the surface $\tx^- -\tx^+ = -2t = \epsilon$.  The explicit terms in the metric \eqref{eq:mxL} then give
\begin{equation}\label{eq:ImK2}
\begin{split}
-\Im \int_{t= - \frac{\epsilon}{2}} \,  d^dx\,\sqrt{h}\, K
&=
	\Im \int_{t= - \frac{\epsilon}{2}}  dx\, d^{d-1} y\, \sqrt{q(y)} \;  \pdv{t} \log \sqrt{\sigma(\tx^+, \tx^-)} \\
&=
	 \frac{\hat m-1}{2} \Im \int_{t= - \frac{\epsilon}{2}}  dx\,\partial_t \log (\tx^+ \tx^-) \times \int d^{d-1}y \,\sqrt{q(y)}  \\
&=
	\frac{\hat m-1}{2} \Im \int_{t= - \frac{\epsilon}{2}}  dx\,\partial_x \left( \log \tx^+ -  \log \tx^-\right) \times A_{\fixM} \\
&=
	2\pi\frac{\hat m-1}{2}A_{\fixM} =  (\hat m-1)\pi A_{\fixM}\,.
\end{split}
\end{equation}
	The factor of $A_{\fixM}$ comes from integrating $\sqrt{q}$ over the longitudinal coordinates $y^I$ as indicated. In passing from the third to the fourth line we have used the fact that $\tx^\pm$ have opposite $i\varepsilon$ pole prescriptions so that their logarithms give oppositely-signed imaginary parts (which then reinforce each other due to the explicit minus sign in the $\tilde x^-$ term in line three).

Summing the two contributions above in \eqref{eq:ImK1} and \eqref{eq:ImK2} gives
\begin{equation}
\label{eq:Cint}
\Im \int_{ \mathscr{C}_\epsilon} d^dx\,  \sqrt{h}\, K = -\pi \,\hat{m}\, A_\fixM.
\end{equation}
But since the full replica geometry involves $n$ copies of $\bulkket$ ket spacetimes and $n$  copies of $\bulkbra$ bra spacetimes, we must include such a semicircle through each in order to form the boundary of a disk around the origin.  Comparing with \eqref{eq:cGB} and setting   $\chi({\mathscr{\tilde U}}_\epsilon)=1$ yields
\begin{equation}\label{eq:RicfixM}
iS_{\fixM} = \frac{i}{16\pi G_N} \int_{\mathscr{U}_\epsilon} \,\eta\sqrt{-g}\, R = -\frac{(n\,\hat m-1)\, A_{\fixM}}{4G_N}\,
= -\frac{(m-1)\, A_{\fixM}}{4G_N}\, .
\end{equation}	

We now conclude our argument by making several observations. The first is simply that comparing \eqref{eq:LHJm} and \eqref{eq:RicfixM} shows -- again in parallel with the Euclidean case -- that for any $n$ the full action $S$ is a Legendre transform of $\check{S}_{m,n}$ with respect to $m$.  So by the usual argument $S$ yields a well-defined variational principle with the `conjugate' boundary conditions still defined by \eqref{eq:mxL} but where we fix $A_{\fixM}$ and instead allow $m$ to vary (still holding $n$ fixed).  In particular, we see that any stationary point of $\check{S}_{m,n}$ will automatically make $S$ stationary with respect to variations of $m$.

Second, we note that we may consider  boundary conditions defined by \eqref{eq:mxL} for fixed $n$ but with {\it both} $A_{\fixM}$ and $m$ free to vary.  Given the observation above, to investigate the status of $S$ for such boundary conditions we need only compute
\begin{equation}
\label{eq:vAfixM}
\fdv{S}{A_{\fixM}} = \fdv{S_{\fixM}}{A_{\fixM}} = i\frac{m-1}{4G_N},
\end{equation}	
where in the first step we have again assumed that we evaluate the result on a stationary point of $\check{S}_{m,n}$.  We see from \eqref{eq:vAfixM} that stationary points of $\check{S}_{m,n}$ with $m=1$ are also stationary points of $S$.  We have thus shown by direct computation that $S$ does indeed yield a well-defined variational principle with boundary conditions defined by \eqref{eq:mxL} and with $n$ fixed but with $m$ free to vary.  And we have also shown that stationarity with respect to $A_{\fixM}$ is equivalent to $m=1$, and thus to $\hat m= 1/n$.  In particular, this is the real-time analogue of the stationarity condition that requires the metric to be smooth in the purely Euclidean context and in any single replica real-time context (i.e., for which $n=1$).

Let us conclude this section with a summary of the requirements for a covering space description of a replica wormhole spacetime to yield a saddle point of our variational principle associated with the full action $S$ for some $n$.  First, it must satisfy the usual Einstein equations away from the timefold surfaces $\homsurfA$, $\homsurfAc$, and $\fixM$.  Second, the extrinsic curvatures at $\homsurfA$, $\homsurfAc$ must be continuous when passing from any bra part of the spacetime to any ket part.  Third, near $\fixM$ it must take the form \eqref{eq:mxL} with $\hat{m} = 1/n$ and with the $i\varepsilon$ prescriptions given above.\footnote{\label{foot:ieps} It would be interesting to attempt to strengthen the above argument and show that the full $\hat{m}=1/n$ `boundary conditions' associated with \eqref{eq:mxL} -- and in particular the $i\varepsilon$ prescription associated with the poles and branch cuts --  are in fact consequences of the equation of motion associated  with stationarity of the action under varying the area of $\fixM$.
In particular, the $i\varepsilon$ prescription is precisely the condition that powers of
$\tx^\pm$ define positive frequency functions.  And as noted in \cref{sec:stateprep} below, for the case where we compute R\'enyi entropies of the vacuum state, we expect this state to enforce boundary conditions on the ket spacetime that in some sense require positive frequency solutions.  Furthermore, since the $i\varepsilon$ prescription is needed only on the past light cone of $\fixM$, it is a UV issue that one expects to be independent of the choice of state.  We thus suspect that a full analysis of the equations of motion and the initial conditions for good quantum states could derive this condition from an entirely real-time point of view.}  Finally, it must be consistent with whatever initial conditions are used to specify the quantum state.  This last requirement will be discussed further in \cref{sec:stateprep}, though we first briefly address the fundamental domain description of our saddles in \cref{sec:fundomain}.

\subsection{The view from a single fundamental domain}
\label{sec:fundomain}

Despite the elegance of the above covering space description,  imposing replica symmetry allows one to reconstruct the full solution from a single fundamental domain.  It is thus often useful to formulate the entire problem in terms of one such domain
$\widehat{\bulk}_n$, containing only a single bra spacetime and a single ket spacetime sewn together along $\homsurfA$ and $\homsurfAc$.  As in \cite{Lewkowycz:2013nqa}, this can be particularly useful for analytically continuing to non-integer $n$ as the parameter $n$ now only appears in the metric through the combination $\hat m = \frac{m}{n}$, which is already allowed to be an arbitrary positive real number.  Below, we also require the spacetime to be invariant under CPT conjugation.

The desired variational principle follows immediately from our discussion above.  We simply define the allowed fundamental domains $\widehat{\bulk}_n$ to be quotients  $\widehat{\bulk}_n = \bulk_n/{\mathbb Z}_n$ of the replica- and conjugation-symmetric covering spaces $\bulk_n$ satisfying the boundary conditions of \cref{sec:realtime}.  We then define the action
\begin{equation}
S^{\text{fund}}_n\left(\widehat{\bulk}_n \right) = \frac{1}{n} S_n\left({\bulk}_n \right) = 2i  \Im \left( S_{\text{gr},n}^k \right) -i \left(\frac{(\hat m - \frac{1}{n})A_{\fixM}}{4G_N} \right),
\end{equation}
where $S_{\text{gr},n}^k$ is the action \eqref{eq:EHact} evaluated on the ket part of  $\widehat{\bulk}_n$, and where in the last step we have used \eqref{eq:RicfixM}.  This Lorentz-signature action is always purely imaginary so that the associated weights in the path integral are real.  In particular, since saddles will again have $\hat m = 1/n$, the R\'enyi entropies \eqref{eq:renyiMn} computed by any given such saddle take the form
\begin{equation}
S^{(n)}_\regA = 2 \frac{n}{n-1} \Im \left(
S_{\text{gr},n}^k - S_{\text{gr},1}^k\right)\, .
\end{equation}	
This is manifestly real, and coincides with the Euclidean answer when analytic continuation can be performed.  This will be verified explicitly in \cite{Colin-Ellerin:2020exa} for the examples studied there.

\section{State preparation}
\label{sec:stateprep}

Thus far we have imagined starting from an initial (perhaps mixed) state $\rho_0$ at some $t_0$ and evolving it to the time $t$ of interest taking into account any  real-time sources between $t_0$ and $t$. By `time' $t$, we in fact mean that we choose some Cauchy surface $\Cbdy$, and similarly for $t_0$. To be definite, we take $t$ to lie to the future of $t_0$.  In the holographic context these will refer to boundary Cauchy surfaces, and we will allow boundary sources which will affect the quantum state of the bulk within their causal future.\footnote{The effects of such sources on the saddle-point geometries used to calculate R\'enyi entropies need not be confined to this causal future.  The point here is that such solutions are not constructed by solving a Cauchy problem with initial data in the past. Instead, they involve timefolded spacetimes with boundary conditions on both sides.  Furthermore, the equations of motion may fail to be hyperbolic in any complex parts of the spacetime.}

While this setting is natural, it raises the question of precisely how the state $\rho_0$ is to be specified.  In the discussion above, we have generally supposed that we have been given the explicit matrix elements of $\rho_0$ in a basis defined by field eigenstates at the time $t_0$.  However, at least in field theoretic contexts, we should admit that it can be difficult to obtain such a description for interesting states.
Let us therefore briefly remark on other methods that can be used to specify $\rho_0$, and which are also readily incorporated into our discussion.

One strategy is to choose a familiar state that allows for a particularly simple treatment and then to construct more complicated states by adding sources between $t_0$ and $t$.  For example, one might take the initial state to be the vacuum, and perhaps also taking $t_0$ to lie in the far past.  This has several advantages. In the free field limit, the vacuum initial conditions require positive-frequency boundary conditions at the initial time, cf., \cite{Marolf:2004fy} for a discussion in language similar to that used here.  Similarly, for vacuum gravity in \AdS{3} the geometry in the far past should be diffeomorphic to global \AdS{3} with all boundary gravitons being of positive frequency. In higher dimensions or when matter fields are coupled to gravity, we still expect an analog of the positive-frequency boundary condition to hold, though making it precise might require employing a suitable `$i\varepsilon$' prescription.\footnote{ In an asymptotically flat spacetime it will suffice to ensure that the linearized gravitational solutions in the far past only have support on positive frequency incoming radiative modes.}

Alternately, it may be useful to consider states that can be prepared by slicing open a Euclidean path integral. One can then implement the past boundary condition by simply attaching this Euclidean path integral.  At the semiclassical level, this will then require the bulk spacetime to satisfy appropriate Euclidean boundary conditions in the far past.   The vacuum can of course be treated in this way, as can the thermofield double state, or deformations of these states by operator insertions in the Euclidean section; see e.g. related discussions in \cite{Faulkner:2017tkh,Botta-Cantcheff:2015sav,Marolf:2017kvq,Haehl:2019fjz}.

It can thus be useful to allow Euclidean sections of an a priori real-time path integral for use in preparing states.  However, we emphasize that this is a not matter of necessity, but is only a matter of expedience.  Such Euclidean sections are a useful technical simplification to enable us exploit known features of the Euclidean path integral to give a geometric picture of the initial condition.   In particular, the use of such Euclidean sections will not restrict in any way the possible presence of non-analytic sources between $t_0 $ to $t$.

A particularly simple example is provided by the gravitational Schwinger-Keldysh geometries discussed in \cite{Glorioso:2018mmw,Chakrabarty:2019aeu,Jana:2020vyx} which capture the real-time evolution of the thermal density matrix of the boundary CFT (see also \cite{vanRees:2009rw}) in the absence of any sources in the Lorentzian evolution. This example is described in  \cref{app:thermal}, where it is used to illustrate some of the general features discussed above.

\section{Discussion}
\label{sec:Disc}

Our work above provides a framework for discussing replica wormholes within a real-time formalism, and in particular in contexts that may include non-analytic sources.  We described a Lorentz-signature bulk gravitational path integral with boundary conditions  associated with computing (swap) R\'enyi entropies in a dual field theory, and which allows configurations with the topology of replica wormholes.  Real Lorentz-signature configurations of this type contain timefolds and a cosmic brane `splitting surface.'   We carefully formulated a variational principle for such spacetimes that yields the Einstein-Hilbert equations of motion away from the timefolds and the splitting surface, and which imposes natural constraints at these surfaces.  In particular, stationarity at the splitting surface forbids having a delta-function contribution to the scalar curvature as defined by a Lorentz-signature (or more generally, complex) generalization of the $2d$ Gauss-Bonnet theorem.  Explicit examples of such real-time replica wormhole saddles will be presented in a companion paper \cite{Colin-Ellerin:2020exa}.    And while we focused here on bulk gravity described by an Einstein-Hilbert action, the generalization to include perturbative higher curvature terms is straightforward using results from appendix B of \cite{Dong:2019piw}.

The fundamental formulation of our real-time path integral involved only real Lorentz-signature configurations.  However,  solutions to the above stationarity conditions are necessarily complex.  Accessing the associated saddles thus requires deforming the original real contour of integration.  This should not be a surprise.
While real-time equations of motion must be real, this need not be true of the boundary conditions imposed by any particular quantum state.   Indeed, instantons that describe gravitational tunneling are famously associated with Euclidean stationary points that again can be accessed only by deforming the original contour of integration specified by a Lorentzian path integral.

Importantly, however,  at least for replica-symmetric saddles (preserving both a ${\mathbb Z}_n$ cyclic symmetry and conjugation symmetry) we found that the metrics to be real in the region spacelike separated from the splitting surface. As a result, contributions of such  spacelike-separated regions to $S$ cancel between the bra and ket parts of the spacetime.  In particular,  so long as they remain spacelike separated from the splitting surface, we can move the bulk timefold along $\homsurfA, \homsurfAc$ forward or backward in time as we please without changing the path integral weight of our replica-wormhole.  This remains true even if we move  $\homsurfA, \homsurfAc$ at the AdS boundary, where such deformations correspond to changing the time $t$ at which our (swap) R\'enyi is computed.  This feature is an important hallmark of unitarity in a dual field theory interpretation, which would indeed require such R\'enyi's to be time-independent.

The above argument suffices to show critical features of unitarity at boundary times that are spacelike separated from the splitting surface, and in contexts where the bulk computation is controlled by a single replica- and conjugation-symmetric saddle.   But it is clearly of interest to understand whether and how bulk replica wormholes implement the expected unitarity more generally.   In particular, since the formal $n\rightarrow 1$ limit of an on-shell splitting surface is an extremal surface, and since extremal surfaces must be spacelike separated from any part of the AdS boundary not causally related to their boundary anchors \cite{Wall:2012uf,Headrick:2014cta}, it is natural to ask whether general on-shell splitting surfaces with any $n>1$ must also be spacelike separated from corresponding regions of AdS boundary.  And it is also clearly important to understand the possible effect of saddles in which replica symmetry is broken.

Now, as a matter of principle, an inherently real-time prescription for constructing saddle-point geometries of gravitational replica path integrals is critical to describing physics in the presence of non-analytic sources.  However, in some contexts it will in practice  be convenient to proceed by studying a related Euclidean problem and analytically continuing the resulting Euclidean saddle.  This may in particular be useful when the initial state can be prepared using a Euclidean path integral and when any sources are analytic functions of time.  In that context, one may imagine computing entropies associated with general Euclidean choices of the region ${\cal A}$ and then analytically continuing parameters to obtain entropies for general Lorentzian regions as in \cite{Almheiri:2019qdq,Hartman:2020swn,Goto:2020wnk}. Analytically continuing the Euclidean replica geometry in this way must give a solution to the variational problem described in \cref{sec:skreplicas}.  Indeed, if the analytic continuation is performed using the prescription described below \eqref{eq:mv} then it is manifest that the result will satisfy the conditions at the real-time splitting surface associated with \eqref{eq:mxL}.  And it is also manifest that analytic continuation of a Euclidean saddle will solve the standard Einstein equations away from $\fixM$.

Of particular interest may be the way that such analytic continuations glue together the bra and ket parts of the resulting Lorentz-signature spacetime.  This gluing is naturally described by a smooth excursion into the space of complex metrics, which then becomes our $i\varepsilon$ prescription in the limit where the curve connecting the Lorentz-signature bra and ket branches becomes very tight.

A particularly clean example was described in \cite{Jana:2020vyx} based on earlier work of \cite{Glorioso:2018mmw} for the case of replica number $n=1$.  The context there involved computing boundary correlation functions of light fields in a thermal state in the limit of vanishing bulk Newton constant $G_N$ with a bulk theory described by pure AdS gravity. As this example may provide inspiration for future work, we recall it briefly in  \cref{app:thermal}, commenting briefly on the extension to general $n$ and emphasizing the way that it displays the reality of the weight $e^{iS}$ and the associated facets of unitarity described above.

Let us now conclude with a few further brief comments on future directions.  First, while it sufficed for our current purposes, there is admittedly something unsatisfying about using an $i\varepsilon$ prescription to define the boundary conditions at $\fixM$ for general $\hat m$. It would thus be interesting to understand whether the allowed configurations could be generalized in some natural way so that our $i\varepsilon$ prescription arose from solving the Einstein equations subject to boundary conditions imposed by natural quantum states.  Indeed, the latter are naturally complex.   One can certainly imagine that such a prescription would follow for vacuum states (as it does in the free field case, see e.g. section 4.3 of \cite{Marolf:2004fy}), and since the $i\varepsilon$ prescription controls a UV singularity on a lightcone one may hope that it in fact follows for more general states due to the requirement that they agree with the vacuum in the UV.

Second, it is clearly of interest to extend our analysis beyond leading order in the bulk Newton constant $G_N$ to include one-loop back-reaction from quantum fields. Here one would like to understand the relationship to classic discussions \cite{Anderson:1986ww,MCD} of back-reaction in the presence of Lorentz-signature topology change.  But the extension to include one-loop back-reaction is also important for many applications to black hole evaporation (where one expects the relevant replicas to be saddles only for the one-loop-corrected effective action \cite{Penington:2019npb,Almheiri:2019psf}. In particular, if one can extend the above discussion of dual field theory unitarity to the bulk one-loop level,  it should provide a bulk argument that black hole evaporation not only yields a Page curve as in \cite{Almheiri:2019qdq,Penington:2019kki} but that it implements fully unitary evolution as expected for a field theory dual.

\acknowledgments
It is a pleasure to thank  Veronika Hubeny, R.~Loganayagam, and Henry Maxfield  for illuminating discussions. We would also like to thank
Amirhossein Tadjini for comments on the draft.
XD, SCE, DM, and MR would like to thank KITP, UCSB for hospitality during the workshop ``Gravitational Holography'' where this work was initiated.
XD was supported in part by the National Science Foundation under Grant No.\ PHY-1820908 and by funds from the University of California.  SCE was  supported by U.S.\ Department of Energy grant {DE-SC0019480} under the HEP-QIS QuantISED program.  DM and ZW were supported by NSF grant PHY1801805 and funds from the University of California. MR  was supported by  U.S. Department of Energy grant DE-SC0009999 and by funds from the University of California. This research was also supported in part by the National Science Foundation Grant No.\ NSF PHY-1748958 to KITP.

\appendix

\section{An alternate accounting scheme}
\label{app:alt}

\begin{figure}[h]
\begin{center}
\begin{tikzpicture}[scale=1.5]
\foreach \x in {0}
{
\draw[red,thick]  (\x+1,0) .. controls (\x+2,-0.3) and (\x+3,+0.5) .. (\x+4,0);
\draw[brown,thick] (\x-1,0) .. controls (\x-2.5,+0.2)  and (\x-3,-0.5) .. (\x-4,-0.25);
\draw[red,thin, dashed]  (\x,0) .. controls (\x+0.5,0.2) and (\x+0.9,+0.1) .. (\x+1,0);
 \draw[brown,thin,dashed] (\x,0) .. controls (\x-0.5,-0.2)  and (\x-0.9,-0.1) .. (\x-1,0);
\draw[black] (\x-2,-2) -- (\x,0) -- (\x+2,-2);
\draw[blue, fill =blue] (\x,0) circle [radius = 3pt];
\draw[thin,dotted,black] (\x-1,0) -- (\x,-1) -- (\x+1,0);
\draw[thick,orange] (\x-1,0) .. controls (\x-0.7,-0.45)  and (\x+0.7,-0.45) .. (\x+1,0);

\draw[thick, orange, -> ] (1,0) -- ++ (0.7,-0.4) node [right]{$n_{\scriptscriptstyle{+ \epsilon}}^{\sf k}$} ;
\draw[thick, orange, -> ] (-1,0) -- ++ (-0.7,-0.4) node [left]{$n_{\scriptscriptstyle{- \epsilon}}^{\sf k}$} ;
\draw[thin, orange, -> ] (1,0) -- ++ (0.7,0.4) node [right]{${n}_{\scriptscriptstyle{+ \epsilon}}^{\sf b}$} ;
\draw[thin, orange, -> ] (-1,0) -- ++ (-0.7,0.4)  node [left]{${n}_{\scriptscriptstyle{- \epsilon}}^{\sf b}$} ;
\foreach \xend in {\x+1, \x-1}
{
	\draw[black,fill=black] (\xend,0) circle [radius = 1.5pt];
}
\node at (-1,0) [above=1pt]{$\color{gray}{\fixM_\epsilon^-}$};
\node at (1,0) [above=1pt]{$\color{gray}{\fixM_\epsilon^+}$};
\draw[ultra thin, blue] (\x-1,0) arc (-180:0:1cm);
\node at (\x+3,0.1) [above=2pt]{$\homsurfA^\epsilon$};
\node at (\x-3,-0.15) [above]{$\homsurfAc^\epsilon$};
\node at (\x,0) [above=2pt] {$\fixM$};

\draw[thick,olive,->] (\x + 3, -1.5) -- ++ (0,1) node[right]{$t$};
\node at (\x+0.5,-0.1){$\scriptstyle{\mathscr{U}_\epsilon}$};
\node at (\x,-0.3) [below]{$\partial\tilde{\mathscr{U}}_\epsilon$};
\node at (\x,-1.3) [below]{$\Cbulk^\epsilon$};
\draw[dotted,black, ->] (\x-2.5,-0.1) ..  controls (\x-2, -1) and (\x-1.3, -1.2) .. (\x-0.2,-1.5);
\draw[dotted,black, ->] (\x+2.7,0.1) ..  controls (\x+2.3, -1) and (\x+1.8, -1.2) .. (\x+0.2,-1.5);
}
\end{tikzpicture}
\caption{A choice of $\partial \tilde{\mathscr{U}}_\epsilon$ (orange curve) whose intersection with $\tilde \Sigma_t$ is not orthogonal. We have indicated its timelike normal at the locus where the regulating surface intersects the bulk Cauchy slice both into the ket part of the spacetime, as well as the corresponding normal from the bra side. The latter has been reflected up into the future to make clear that in the limit of the smooth join the inner product is between anti-parallel vectors owing to the opposite time-orientation on $\bulkbra$ relative to that on $\bulkket$.  }
\label{fig:epsregulator2}
\end{center}
\end{figure}
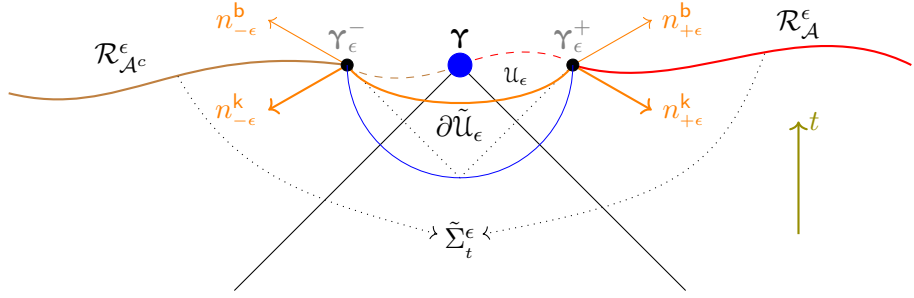

As described at the end of  \cref{sec:cover}, it is sometimes useful to absorb the brane contribution $S_{\fixM}$ into the contributions from the ket and bra parts of the spacetime. For instance, one can redefine the terms in \eqref{eq:TotalS}  to
\begin{equation}
\label{eq:tildeS}
S = \left(S^k_\text{gr} + \frac{1}{2} S_\fixM\right)-  \left(S^b_\text{gr} -\frac{1}{2} S_\fixM\right) \equiv  \tilde S^k_\text{gr} - \tilde S^b_\text{gr}\,.
\end{equation}
In order that \eqref{eq:tildeS} agree precisely with \eqref{eq:TotalS}, we take
$\tilde S_\text{gr}^k = \lim_{\epsilon \rightarrow 0} \tilde S_\text{gr}^{k, \epsilon}$ with
\begin{equation}\label{eq:EHactFD}
\begin{split}
\tilde S_\text{gr}^{k, \epsilon}
&=
	\frac{1}{16\pi G_N} \int_{\bulkket_\epsilon}\, d^{d+1} x\, \sqrt{-g} \left[ R + \frac{d(d-1)}{\lads^2} \right]
	+ S_\text{bdy}^\epsilon + S_\text{corner}^\epsilon \,, \\
S_\text{bdy}^\epsilon
&=
	\frac{1}{8\pi G_N} \int_{\bdy}\, d^{d} x\, \sqrt{|\gamma|} \, K  + \frac{1}{8\pi G_N} \int_{\Cbulk^\epsilon \cup \partial\mathscr{U}_\epsilon} \,  \sqrt{|h|} \, K
	\\
S_\text{corner}^\epsilon
&=
	\frac{1}{16\pi G_N} \int_{\fixM_\epsilon^+ \cup \fixM^-_\epsilon}\, dA\, \cosh^{-1}\left( n_\epsilon^{\sf k} \cdot n_\epsilon^{\sf b}
	\right) \,,
\end{split}	
\end{equation}	
with a complex-conjugate prescription for $\tilde S_\text{gr}^b$. Here $\gamma_{\mu\nu}$ is the induced metric on the boundary $\bdy$ and $h_{ij}$ that on the interior boundary. $K$ is the trace of the extrinsic curvature defined using the appropriate normal.  We have also allowed for possible `corner' terms associated with choosing $\partial {\mathscr{U}}_\epsilon$ to have non-orthogonal intersection with the bulk timefold $\tilde \Sigma_t$ as shown in \cref{fig:epsregulator2} (the orange curve).

We consider the case where the regulating surface $\partial \mathscr{U}_\epsilon$ is everywhere spacelike.  By  doing so we no longer pick up the poles in the boost angle at the past light cone. The contribution \eqref{eq:ImK1} now instead can be attributed to this corner term (cf., for instance \cite{Neiman:2013ap} where this was discussed earlier). The explicit expression for the corner contribution is given in terms of the inner product (or relative boost) between the unit normals $n_\epsilon^{\sf k}$ and $n_\epsilon^{\sf b}$  to the regulator  surfaces $\mathscr{U}_\epsilon$ on the ket and the bra geometries, respectively. Though we integrate the corner contribution over the regulated codimension-2 fixed point loci $\fixM_\epsilon^\pm$, the essential contribution can be understood from the $1+1$ dimensional example, where each corner contributes $\frac{i}{16 G_N}$ to $\tilde{S}^{k,\epsilon}_\text{gr}$.  We could also consider a more general non-orthogonal intersection as mentioned in \cref{fn:generalcorner}, where the final result  works out albeit with a slightly modified accounting of the contributions.

The boundary terms and the corner terms are written out here in the standard metric formulation above. The corner terms were derived in \cite{Jubb:2016qzt}  using the differential formulation of the action (using a non-holonomic tetrad basis) which has the added advantage of seeing quite explicitly that they are essential to having a well-defined variational principle.  As in \cref{sec:cover}, the extrinsic curvature terms
$S_\text{bdy}^\epsilon$ receive subtle contributions when the boundary transitions from being spacelike to timelike, though these again have a simple universal form in the differential formalism \cite{Jubb:2016qzt}.

\section{Smoothing the bra-ket gluing via excursions into the complex plane}
\label{app:thermal}

In this appendix  we recall a particularly clean example from \cite{Glorioso:2018mmw,Chakrabarty:2019aeu,Jana:2020vyx} of the gluing of the bra and ket branches of a saddle-point.  In this case the gluing is performed by making a smooth excursion into the space of complex metrics, which smooths out both the timefolds and the cosmic brane splitting surface described in the main text.  As discussed in the aforementioned references, the fact that the construction solves the field equations with the correct boundary conditions is critical for obtaining the proper physics for boundary correlators.  This example also gives a very explicit illustration of the reality of the weight $e^{iS}$ associated with replica-symmetric saddles.

\begin{figure}[htbp]
\begin{center}
\begin{tikzpicture}
\foreach \x in {0, 4,6,8}
{
	\draw[color=orange,thick,-<-]  (\x,0) .. controls (\x -1.3, -1.5) and (\x+2.7,-0.9) .. (\x+1,0.3);
	\draw[color=blue,thick, ->-] (\x,0)  -- ++ (0,4);
	\draw[color=blue,thick, -<-] (\x+1,0.3) --  ++ (0,3.7);
	\draw[color=blue,fill=blue] (\x,0) circle (1pt);
	\draw[color=blue,fill=blue] (\x+1,0.3) circle (1pt);
	\node at (\x+0.5,-0.3) [below]{$\rho_0$};
}
\node at (0,2)  [left]{$\mathcal{U}(t;t_0)$};
\node at (1,2) [right=2pt]{$\mathcal{U}(t;t_0)^\dagger$};
\foreach\x in {5,7}
{
	\draw[color=blue,thick] (\x,4) .. controls (\x+0.3,4.3) and (\x+0.6,4.3) .. (\x+1,4);
}
\draw[color=blue,thick] (4,4) .. controls (5.67, 5) and (7.33,5) .. (9,4);
\end{tikzpicture}
\caption{The thermal density matrix of a field theory can be prepared from a Euclidean path integral on a $\bdy = {\bf S}^1_{t_{_\text{E}}} \times \Cbdy$ with $t_{_\text{E}} = \beta + i t$.  Its time evolution $\rho(t)$ in real-time is indicated on the left. On the right we depict the gluing of three copies of this real-time density matrix to compute the third moment $\Tr(\rho(t)^3)$. Since there are no sources in the Lorentzian evolution the forward backward evolution legs of this real-time contour cancel pairwise leaving behind a path integral that is localized on the Euclidean section, now on a thermal circle that it three times larger. This is a consequence of the standard collapse rules of real-time path integral contours.  }
\label{fig:bdy3thermal}
\end{center}
\end{figure}
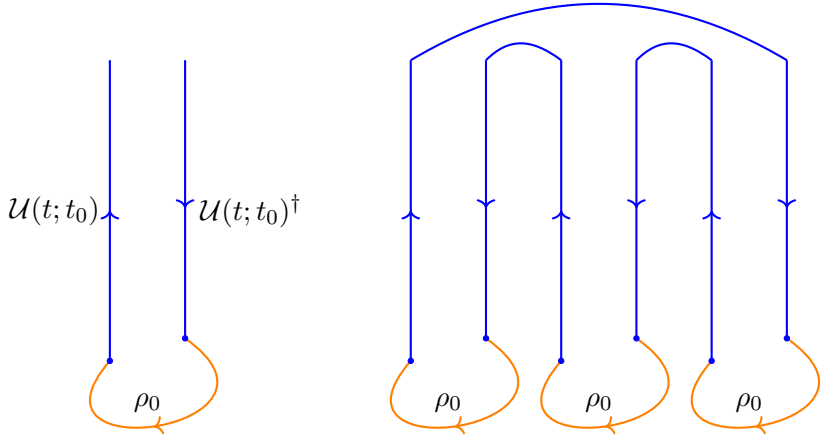

We will borrow from the example discussed in \cite{Glorioso:2018mmw,Chakrabarty:2019aeu,Jana:2020vyx}  which corresponds to the case of replica number $n=1$.  The context there involved computing boundary correlation functions of light fields in a thermal state in the limit of vanishing bulk Newton constant $G_N$. As a result, back-reaction from the quantum fields could be neglected.  The problem thus reduced to studying quantum fields on a fixed bulk background that (for $n=1$) was just the $(d+1)$-dimensional AdS-Schwarzschild black hole with specified inverse temperature $\beta$.  For a thermal problem, one may think of starting with the Euclidean solution, constructing the full complex black hole geometry by analytic continuation, and then choosing to deform the original Euclidean slice of this complex geometry  as desired into an arbitrary contour of real bulk dimension $d+1$.  For the purposes of studying boundary correlation functions of light fields at general Lorentzian times $t> 0$, it was useful in \cite{Jana:2020vyx} to take the resulting slice at the AdS boundary to extend to $t=+\infty$ along the positive real Lorentzian axis as shown in the left panel of \cref{fig:bdy3thermal}.  The right panel shows the generalization of the boundary geometry to $n=3$.  In both cases, we can think of the boundary conditions as forcing $t$ to follow a certain contour in the complex time-plane.

%

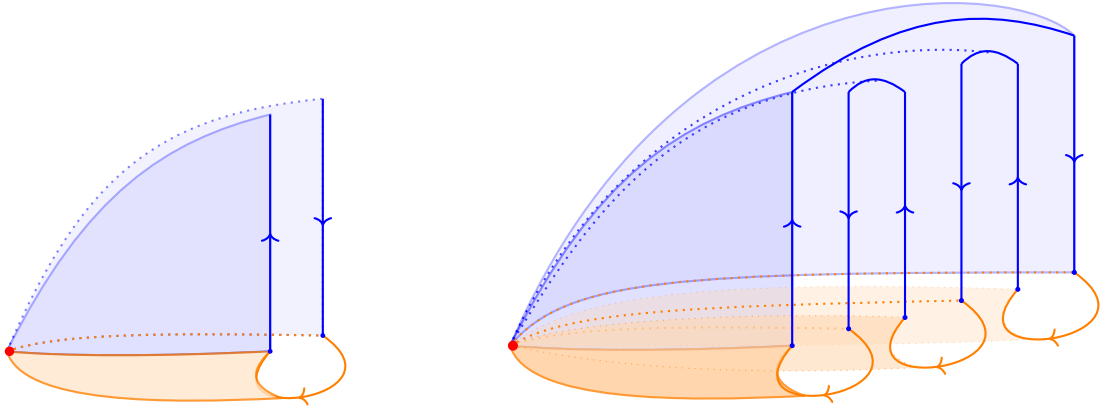
\begin{figure}[htbp]
\begin{center}
\begin{tikzpicture}[scale=0.7]
\foreach \x in {0}
{
	\draw[color=orange,thick,-<-]  (\x,0) .. controls (\x -1.3, -1.5) and (\x+2.7,-0.9) .. (\x+1,0.3);

	\filldraw[color=blue, fill=blue!20,opacity=0.6,thick] (\x-5,0) .. controls (\x-3.5,3) and (\x-2,4) .. (\x,4.5) -- (\x,0) -- (\x,0) .. controls (\x - 2,-0.1) and (\x-4, -0.1) .. (\x-5,0);

	\filldraw[color=orange, fill=orange!20,opacity=0.8,thick] (\x+0.22,-0.89) .. controls (\x - 2,-1) and (\x-4.7, -1) .. (\x-5,0)
	--  (\x-5,0) .. controls (\x - 4,-0.1) and (\x-2, -0.1) .. (\x,0) -- (\x,0) .. controls (\x-0.2,-0.2) and (\x-0.58,-0.6) .. (\x+0.22,-0.89);

	\filldraw[color=blue,fill=blue!10,opacity=0.5,thick,dotted] (\x-5,0) .. controls (\x-3.5,3.5) and (\x-2,4.5) .. (\x+1,4.8) -- (\x+1,0.3) --
		(\x+1,0.3) .. controls (\x-2, 0.35) and  (\x - 4,0.35)  .. (\x-5,0);

		\draw[color=blue,thick, ->-] (\x,0)  -- ++ (0,4.5);
	\draw[color=blue,thick, -<-] (\x+1,0.3) --  ++ (0,4.5);
	\draw[color=blue,fill=blue] (\x,0) circle (1pt);
	\draw[color=blue,fill=blue] (\x+1,0.3) circle (1pt);
	
	\draw[color=orange,thick,dotted] (\x-5,0) .. controls (\x - 4,0.35) and (\x-2, 0.35) .. (\x+1,0.3);

	\draw[color=red,fill=red,thick] (\x-4.95,0) circle (2pt);
}
\end{tikzpicture}
\hspace{1cm}
\begin{tikzpicture}[scale=0.75]
\coordinate (ai) at (0,0);
\coordinate (af) at (1,0.3);
\coordinate (bi) at (2,0.5);
\coordinate (bf) at (3,0.8);
\coordinate (ci) at (4,1);
\coordinate (cf) at (5,1.3);
\coordinate (off) at (0.22,-0.89);

 	\filldraw[color=blue,fill=blue!20,opacity=0.6,thick] ($(ai) + (-5,0)$) .. controls ($(ai) + (-3.5,3)$) and ($(ai) + (-2,4)$) .. ($(ai)  +(0,4.5)$)
		-- (ai) --  (ai)  .. controls ($(ai) + (-2,-0.1)$) and ($(ai) + (-4,-0.1)$) .. ($(ai) + (-5,0)$) ;

	\draw[color=blue,thick,dotted] ($(ai) + (-5,0)$) .. controls ($(ai) + (-3,4.5)$) and ($(ai) + (1,5.5)$) .. ($(ci)  +(-0.5,4.2)$) ;

	\draw[color=blue,thick,dotted] ($(ai) + (-5,0)$) .. controls ($(ai) + (-3.5,2.5)$) and ($(ai) + (-2,4.5)$) .. ($(bi)  +(-0.5,4.2)$)  ;

	\filldraw[color=blue,fill=blue!20,opacity=0.3,thick]
		 ($(ai) + (-5,0)$) .. controls ($(ai) + (-2,7)$) and ($(bi) + (2,6)$) .. ($(cf)  +(0,4.2)$) -- (cf) -- (cf) ..  controls ($(ai) + (-2,1.3)$)  and  ($(ai) + (-4,1.2)$) ..  ($(ai) + (-5,0)$) ;
		
 	\filldraw[orange, fill=orange!40,opacity=0.8,thick] ($(ai)+(0.22,-0.89)$) .. controls ($(ai) + (-2,-1)$) and ($(ai) + (-4.7, -1)$) ..
 		($(ai) + (-5,0)$) -- ($(ai) + (-5,0)$) .. controls ($(ai) + (-4,-0.1)$) and ($(ai) + (-2,-0.1)$) .. (ai) --
 		(ai) .. controls ($(ai) +(-0.2,-0.2)$) and ($(ai)+(-0.58,-0.6)$) .. ($(ai)+(0.22,-0.89)$) ;
	\draw[color=orange,thick,dotted]  ($(ai) + (-5,0)$) .. controls  ($(ai) + (-4,0.35)$) and ($(ai) + (-2,0.35)$) .. (af);
	
  	\filldraw[orange, fill=orange!30,opacity=0.35,thin,dotted]
		 ($(ci)+(0.22,-0.89)$) .. controls ($(ci) + (-2,-1)$) and ($(ci) + (-4.7, -1)$) .. ($(ai) + (-5,0)$) --
		 ($(ai) + (-5,0)$) .. controls ($(ai) + (-4,0.8)$) and ($(ai) + (-2,1.2)$) .. (ci) --
		 (ci) .. controls ($(ci) +(-0.2,-0.2)$) and ($(ci)+(-0.59,-0.65)$) .. ($(ci)+(0.22,-0.89)$) ;
	\draw[color=orange,thick,dotted]  ($(ai) + (-5,0)$) .. controls  ($(ai) + (-4,1.2)$) and ($(ai) + (-2,1.3)$) .. (cf);

	\filldraw[orange, fill=orange!30,opacity=0.6,thin,dotted]
		($(bi)+(0.22,-0.89)$) .. controls ($(bi) + (-2,-1)$) and ($(bi) + (-4.7, -1)$) .. ($(ai) + (-5,0)$) --
		($(ai) + (-5,0)$) .. controls ($(ai) + (-4,0.5)$) and ($(bi) + (-2,0.1)$) .. (bi) --
 		(bi) .. controls ($(bi) +(-0.2,-0.2)$) and ($(bi)+(-0.59,-0.65)$) .. ($(bi)+(0.22,-0.89)$) ;
 	\draw[color=orange,thick,dotted]  ($(ai) + (-5,0)$) .. controls  ($(ai) + (-4,0.6)$) and ($(ai) + (-2,0.7)$) .. (bf);

	\draw[color=orange,thick,-<-]  (ai) .. controls ($(ai) + (-1.3, -1.5)$) and ($(ai) +(2.7,-0.9)$) .. (af);
	\draw[color=blue,thick, ->-] (ai)  -- ++ (0,4.5);
	\draw[color=blue,thick, -<-] (af) --  ++ (0,4.2);
	\draw[color=blue,fill=blue] (ai) circle (1pt);
	\draw[color=blue,fill=blue] (af) circle (1pt);

	\draw[color=orange,thick,-<-]  (bi) .. controls ($(bi) + (-1.3, -1.5)$) and ($(bi) +(2.7,-0.9)$) .. (bf);
	\draw[color=blue,thick, ->-] (bi)  -- ++ (0,4);
	\draw[color=blue,thick, -<-] (bf) --  ++ (0,4.2);
	\draw[color=blue,fill=blue] (bi) circle (1pt);
	\draw[color=blue,fill=blue] (bf) circle (1pt);

	\draw[color=orange,thick,-<-]  (ci) .. controls ($(ci) + (-1.3, -1.5)$) and ($(ci) +(2.7,-0.9)$) .. (cf);
	\draw[color=blue,thick, ->-] (ci)  -- ++ (0,4);
	\draw[color=blue,thick, -<-] (cf) --  ++ (0,4.2);
	\draw[color=blue,fill=blue] (ci) circle (1pt);
	\draw[color=blue,fill=blue] (cf) circle (1pt);

	\draw[color=blue,thick] ($(af) + (0,4.2)$) .. controls ($(af)  + (0.3,4.5)$) and ($(af)  + (0.6,4.5)$) .. ($(bi)+(0,4)$);
	\draw[color=blue,thick] ($(bf) + (0,4.2)$) .. controls ($(bf)  + (0.3,4.5)$) and ($(bf)  + (0.6,4.5)$) .. ($(ci)+(0,4)$);
	\draw[color=blue,thick] ($(ai) + (0,4.5)$) .. controls ($(bi)  + (0,5.5)$) and ($(ci)  + (-0.5,5)$) .. ($(cf)+(0,4.2)$);

	\draw[color=red,fill=red,thick] ($(ai) + (-4.95,0)$)circle (2pt);

\end{tikzpicture}
\caption{ Gravity dual of the field theory computation illustrated in \cref{fig:bdy3thermal}. The bulk thermal density matrix $\rho(t)$ is obtained by slicing open the Euclidean black hole spacetime, with suitable real-time evolution ending on the cuts thus created. The cuts are pinned at the tip of the Euclidean cigar, which corresponds to the bifurcation surface of the black hole. Lorentzian sections are the time-evolution of the initial state, which geometrically gives rise to the domain of outer communication This is a particularly convenient choice.  One can also include part of the real-time black hole interior. This does not matter as the unitary evolution cancels between the bra and ket pieces as described. For our choice we have two such copies one for the ket and another for the bra as indicated on the left panel.  On the right we illustrate the ansatz for the computation of $\Tr(\rho(t)^3)$; we have copies of the density matrix constructed gravitationally sewn together capturing the kinematic data. Dynamics dictated by the gravitational equations of motion, will ensure the absence of any singularity at the splitting surface $\fixM$ as discussed in the text.
}
\label{fig:bulk3thermal}
\end{center}
\end{figure}

If we are interested only in the computation of the moments $\Tr(\rho(t)^n)$, then the real-time evolution pieces between the bras and kets of neighbouring copies of $\rho(t)$ cancel pairwise using unitarity of the evolution in the form  $\mathcal{U}(t,0) \mathcal{U}^\dagger(t,0) =1$.  This leaves behind a path integral contour that is localized on the  Euclidean section alone. However, now the Euclidean thermal circle is $n$-times larger, and indeed we recover the fact that the moments of the thermal density matrix are simply partition functions at a rescaled temperature.\footnote{Note that these statements are a simple consequence of unitary evolution and an important consistency requirement of the real-time path integral contours. In Schwinger-Keldysh computations they are usually referred to as collapse rules, see \cite{Haehl:2016pec, Haehl:2017qfl} for a general discussion and implications in the context  of Schwinger-Keldysh and out-of-time-order observables.}

Let us now turn to the holographic description of this example. We first consider the saddle-point spacetime that computes $\Tr \,\rho(t)$ as  described in the papers referenced above.  The above cancellations imply that it is related to the Euclidean AdS-Schwarzschild black hole.  However, in order to allow the insertion of operators at non-zero real times $t$, it is useful to describe the saddle using a different contour through the complexified AdS-Schwarzschild spacetime that also has real Lorentz-signature pieces.

Indeed, a key part of the geometry of \cite{Jana:2020vyx} may represented by a two-sheeted Lorentz-signature spacetime where the sheets are respectively associated with the bra and ket parts of $\rho(t)$. Rather than terminate the Lorentzian evolution at the time $t$ of interest along the boundary,   we choose to extend the spacetime to future timelike infinity along both the ket and the bra segments. In the bulk, we similarly take the bra and ket spacetimes to be sewn to each other long the future event horizon, so that the Lorentzian part of each sheet corresponds to the $t>0$ part of the AdS-Schwarzschild domain of outer-communication; see \cref{fig:bulk3thermal} (Figure 2 of \cite{Jana:2020vyx}). As we discuss below, this allows for a simple presentation of the bra-to-ket sewing using a complexified coordinate chart. This two-sheeted Lorentzian geometry is then glued at $t=0$ onto the Euclidean black hole solution, which is the familiar Gibbons-Hawking cigar spacetime of \cite{Gibbons:1976ue}.

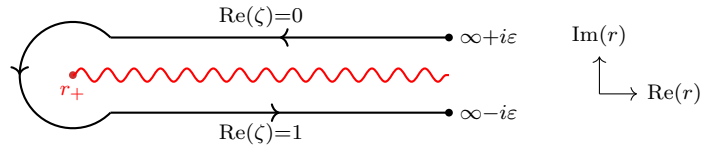
\begin{figure}[h!]
\begin{center}
\begin{tikzpicture}[scale=0.5]
\draw[thick,color=rust,fill=rust] (-5,0) circle (0.45ex);
\draw[thick,color=black,fill=black] (5,1) circle (0.45ex);
\draw[thick,color=black,fill=black] (5,-1) circle (0.45ex);
\draw[ thick,snake it, color=red] (-5,0) node [below] {$\scriptstyle{r_+}$} -- (5,0);
\draw[thick,color=black, ->-] (5,1)  node [right] {$\scriptstyle{\infty+i\varepsilon}$} -- (0,1) node [above] {$\scriptstyle{\Re(\zeta) =0}$} -- (-4,1);
\draw[thick,color=black,->-] (-4,-1) -- (0,-1) node [below] {$\scriptstyle{\Re(\zeta) =1}$} -- (5,-1) node [right] {$\scriptstyle{\infty-i\varepsilon}$};
\draw[thick,color=black,->-] (-4,1) arc (45:315:1.414);
\draw[thin, color=black,  ->] (9,-0.5) -- (9,0.5) node [above] {$\scriptstyle{\Im(r)}$};
\draw[thin, color=black,  ->] (9,-0.5) -- (10,-0.5) node [right] {$\scriptstyle{\Re(r)}$};
\end{tikzpicture}
\caption{ The complex $r$ plane with the locations of the two boundaries and the horizon marked. The grSK contour is a codimension-1 surface in this plane (drawn at fixed $v$). The direction of the contour is as indicated counter-clockwise encircling the branch point at the horizon.}
\label{fig:mockt}
\end{center}
\end{figure}

The sewing of bra to ket along the future horizon can be performed cleanly and explicitly using ingoing Eddington-Finkelstein coordinates (which are regular in this part of the complexified geometry). We promote the radial coordinate $r$ to be complex and take the Lorentzian bra and ket spacetimes to lie on the edges of a branch cut along the positive real $r$-axis.  We then sew the branches together by choosing a curve in the complex $r$-plane that circles the branch point at $r=r_+$ as shown in  \cref{fig:mockt}.  This is conveniently described by introducing a new coordinate $\zeta$ that is sensitive to the branch cut, taking values in $1 +i{\mathbb R}$ along the ket piece and values $0+i{\mathbb R}$ along the bra piece.

Focusing on the thermal state defined by the Rindler patch of \AdS{2}
for simplicity\footnote{ In higher dimensions we have for the Schwarzschild-\AdS{d+1} geometries from \cite{Jana:2020vyx}
\begin{equation}
\begin{split}
	ds^2 &= -r(\zeta)^2\, f(r)\, dt^2 + i\, \beta\, r(\zeta)^2 \, dt \, d\zeta + r^2(\zeta)\, d\vb{x}_{d-1}^2 \,, \qquad f(r) = 1-\frac{r_+^d}{r^d} \,,\\
	\zeta &= \frac{i\,d}{2\pi(d-1)}\,\left( \frac{r}{r_+} \right)^{d-1}\,
	{}_2F_1\left(1, 1-\frac{1}{d}, 2-\frac{1}{d};\, \frac{r^d}{r_+^d}\right) \,.
\end{split}\end{equation}
where the branch-cut of the hypergeometric function has been placed to run from $r_+$ to $\infty$.}
we can write the spacetime metric for the Lorentzian part of the evolution as
\begin{equation}\label{eq:skgrav2}
ds^2 =  - r_+^2\,\csc^2(\pi \zeta)\, \left( dt^2+ \frac{2\pi i}{ r_+}\, dt\, d\zeta\right)\,,
\end{equation}	
where we have chosen $\zeta = \frac{1}{2\pi i}\, \log \left(\frac{r-r_+}{r+r_+}\right)$ which makes explicit the placement of the aforementioned logarithmic branch cut.  On the individual ket and bra copies we could revert to the standard Rindler coordinates and write the metric as

\begin{equation}\label{eq:rindlerads2}
ds^2 = -(r^2-r_+^2) \, dt^2 + 2\, dt \,dr \,.
\end{equation}	
The metric here is written in ingoing coordinates which is convenient for gluing the bra and ket copies across the future horizon.

The fact that geometry is a saddle point to the Einstein-Hilbert action follows immediately from the fact that the local equations of motion are satisfied everywhere.  In particular, the smoothness of the gluing ensures that the space has no (complex) conical singularities.   In \eqref{eq:skgrav2} we have exploited a set of complex coordinates to go around the location of the cosmic brane. The gluing  along the future horizons is essentially passing from one sheet to another smoothly.\footnote{Note that the passage to the complex coordinates enables us to see smoothness explicitly.  One can also choose to work with real sections gluing the bra and ket copies in a replica symmetric manner along the future horizon. This is the viewpoint originally advocated in \cite{vanRees:2009rw}.}  We also make use of the fact that, while our coordinates $(v,r)$ are not in fact smooth at $\fixM$ (the bifurcation surface), one expects (as motivated in  \cite{Jana:2020vyx})  that the complexified domains of outer communication described by  \cref{fig:mockt} can be smoothly glued onto the Euclidean cigar. Intuitively, one imagines the Lorentzian evolution  on the ket  (bra) part starting (terminating)  at the $t_{_\text{E}} = 0- \varepsilon$  ($t_{_\text{E}} = 0 + \varepsilon$). The Euclidean cigar smoothly connects up these initial and final Cauchy slices. However, as written our coordinate chart  \eqref{eq:skgrav2} does not explicitly include the Euclidean section -- it would be useful to tighten the argument to exhibit the entire spacetime including the initial state preparation as a complex curve in the complex $(v,r)$ space.

To compute the on-shell action we follow the contour integral prescription of \cite{Jana:2020vyx} for the Lorentzian part ($t>0$). One should also include the contribution from the initial state $\rho_0$, which, as we know, is just the Gibbons-Hawking computation of the black hole entropy and  follows from the Euclidean path integral on the cigar. We will include this in the final answer.

Focusing for the present on the real-time section (defined using the above complex regulator), we upgrade the Einstein-Hilbert action to
\begin{equation}\label{eq:skEH}
S_\text{gr} =
	\frac{1}{16\pi G_N} \left[ \oint_{\mathcal{C}} \, d\zeta\, \int \,d^dx\, \sqrt{-g} \, R + 2 \oint_{\mathcal{C}} \, d\zeta\, \int\, d^dx\, \sqrt{|h|} \, \left(K   - 2(d-1) - \frac{1}{d-2}\, {}^h R\right)
	\right]
\end{equation}	
where we have separated out the radial integral from the other directions and have explicitly included the boundary counter-terms (the boundary cosmological constant and the Einstein-Hilbert term). The contour $\mathcal{C}$ we integrate over runs counterclockwise from the bra boundary to the ket boundary encircling the branch point at the horizon, see \cref{fig:mockt}. The contour integral picks up the discontinuity across the cut. However, for the metric functions in \eqref{eq:skgrav2} there is none as the metric enjoys periodicity under $\zeta \to \zeta +1$. So the on-shell action receives no contribution from the Lorentzian sections and collapses completely onto the initial state.\footnote{This cancellation would be obstructed if we turn on sources in the Lorentzian evolution as can be seen from the correlation function calculations in \cite{Jana:2020vyx}. }

Since the initial state path integral coincides with the usual thermal path integral, we correctly obtain  the usual thermal partition sum $\Tr(e^{-\beta H}) = e^{-I_1}$
\begin{equation}\label{eq:thermalTr}
\log \Tr(\rho(t)) = \log \Tr(\rho_0) = \log \Tr(e^{-\beta H})  = 4\pi\, c_\text{eff} \,\text{V}_{d-1}\, \left(\frac{4\pi}{d\,\beta}\right)^d\,
\end{equation}	
 as indeed expected from the Schwinger-Keldysh collapse rules (note that $\beta = \frac{4\pi}{dr_+}$ for the Schwarzschild-\AdS{d+1} black hole).

This construction readily generalizes to the computation of the replica geometries $\bulk_n$ which compute spectral moments of the thermal density matrix for the entire CFT in real-time. We simply lay out $n$-copies of the bra and ket geometries described above (with $r_+$ arbitrary) and glue them in a replica symmetric manner as shown at right in \cref{fig:bulk3thermal}.  This describes the kinematics of the construction as in the main text -- the explicit geometry will be determined by the imposing the Einstein-Hilbert dynamics. In particular, the parameter $r_+$ should chosen to make the resulting complex geometry smooth.

Heuristically, the replica spacetime including the Lorentzian sections should resemble the following. The ket and bra segments of the geometry are confined to $t\geq 0$ and $r\geq r_+$, respectively and the homology surface is the $t=0$ (ingoing slice) with the homology wedge being the future half of domain of outer communication. The cyclic gluing of the replica copies should be  accompanied by suitable smoothing out of the seams along the future horizons by excursions into the complex domain.  The non-trivial part is again localized on the splitting surface, which here is the bifurcation surface $r=r_+$. While the full solution is not yet available in closed form, one can argue how the computation of the on-shell action works.

In the current example,  the variational problem is completely specified by \eqref{eq:EHact}. The only boundary terms are those that are needed for the usual AdS asymptotics which we included in \eqref{eq:skEH}. Now we simply upgrade the contour integral prescription to cover the $n$-fold replica geometry. The crucial point is once again the observation that  the on-shell action receives no contribution from the Lorentzian sections and collapses completely onto the initial state. For the  R\'enyi  entropies one will find pairwise cancellations between bra and ket pieces leaving behind $n$-copies of the thermal boundary conditions, making it clear that $r_+$ should take the value appropriate to a black hole of inverse temperature $n\beta$.  This then results correctly in the $n^\text{th}$ moment of the thermal density matrix giving
\begin{equation}\label{eq:thermalTr2}
e^{-I_n} = \log \Tr(\rho(t)^n) = \log \Tr(\rho_0^n) = \log \Tr(e^{-n\, \beta H})  = 4\pi\, c_\text{eff} \,\text{V}_{d-1}\, \left(\frac{4\pi}{n\, \beta \, d}\right)^d.
\end{equation}	
%


\providecommand{\href}[2]{#2}\begingroup\raggedright\endgroup

\end{document}